\documentclass[prl,longbibliography,showpacs,twocolumn,superscriptaddress,amsmath,amssymb,verbatim]{revtex4-1}
\usepackage{amsmath,amssymb,amsfonts,stmaryrd,wasysym,graphicx,multirow,color,textcomp,subfigure}
\usepackage{url}
\usepackage[colorlinks=true, urlcolor=blue, citecolor=cyan, pdfborder={0 0 0}]{hyperref}

\usepackage{color,soul} 
\usepackage{bm}
\usepackage[normalem]{ulem}

\normalfont
\def\mk{\mathbf{k}}

\newcommand{\tcr}[1]{\textcolor{red}{#1}}

\newcommand{\beginsupplement}{
	\setcounter{table}{0}
	\renewcommand{\thetable}{S\arabic{table}}
	\setcounter{figure}{0}
	\renewcommand{\thefigure}{S\arabic{figure}}
	\setcounter{equation}{0}
	\renewcommand{\theequation}{S\arabic{equation}}
}

\begin{document}
\title{Geometric Superconductivity in 3D Hofstadter Butterfly}

\author{Moon Jip Park}
\email{moonjippark@kaist.ac.kr}
\affiliation{Department of Physics, KAIST, Daejeon 34141, Republic of Korea}

\author{Yong Baek Kim}
\email{ybkim@physics.utoronto.ca}
\affiliation{Department of Physics, University of Toronto, Toronto, Ontario M5S 1A7, Canada}
\affiliation{School of Physics, Korea Institute for Advanced Study, Seoul 02455, Korea}

\author{SungBin Lee}
\email{sungbin@kaist.ac.kr}
\affiliation{Department of Physics, KAIST, Daejeon 34141, Republic of Korea}

\maketitle

{\bf 
	Electrons on the lattice subject to a strong magnetic field exhibit the fractal spectrum of electrons, which is known as the Hofstadter butterfly. In this work, we investigate unconventional superconductivity in a three-dimensional Hofstadter butterfly system. While it is generally difficult to achieve the Hofstadter regime, we show that the quasi-two-dimensional materials with a tilted magnetic field produce the large-scale superlattices, which generate the Hofstadter butterfly even at the moderate magnetic field strength. We first show that the van-Hove singularities of the butterfly flat bands greatly elevate the superconducting critical temperature, offering a new mechanism of field-enhanced superconductivity. Furthermore, we demonstrate that the quantum geometry of the Landau mini-bands plays a crucial role in the description of the superconductivity, which is shown to be clearly distinct from the conventional superconductors. Finally, we discuss the relevance of our results to the recently discovered re-entrant superconductivity of $\textrm{UTe}_2$ in strong magnetic fields.

}
 
The pairing instability of the conventional superconductivity is dictated by the Bloch electronic states near the Fermi surface. These states are described by the semiclassical equations of the motion\cite{gor1959microscopic,eilenberger1968transformation,PhysRevLett.56.2732,MARKIEWICZ19971179,PhysRevX.9.021025,PhysRevLett.16.996}, since the wave functions are well-extended over many lattice sites. In contrast, the wave packets of the electrons subject to a magnetic field are localized with the length scale of the magnetic length. Especially, if the magnetic field is strong enough such that the magnetic length becomes comparable to the lattice length scale, the fractal patterns of the electron spectrum emerge, known as the Hofstadter butterfly\cite{PhysRevB.14.2239}. The electronic states associated with the Hofstadter butterfly have no classical analogs, thus the conventional semiclassical theories are inapplicable. In this extreme quantum limit, what kinds of superconductivity would arise?

Recent experiments shed light on this issue and offer the possible candidate materials for unconventional superconductivity in strong magnetic fields. A prominent example is the discovery of the field-boosted superconductivity in $\textrm{UTe}_2$\cite{Ran684,doi:10.7566/JPSJ.88.043702,doi:10.7566/JPSJ.88.063706,Ran2019,PhysRevB.100.140502,doi:10.7566/JPSJ.88.073701,PhysRevLett.124.086601,PhysRevB.100.220504,Jiao2020,PhysRevLett.123.217001,Braithwaite2019,PhysRevB.101.140503,bae2019anomalous}. The re-entrant superconducting phase occurs in the presence of the tilted field and sustains the field strength up to $60$T\cite{Ran2019}. The corresponding magnetic length roughly corresponds to $\sim 3.3$nm that is comparable to the lattice length scales($a,b,c=4.1,\, 6.0,\, 13.8 \, \AA $)\cite{hutanu2019crystal}. Hence, the theoretical description of such re-entrant superconductivity would be insufficient without the full consideration of the Landau quantization as well as the lattice effect. In addition, there are other superconductors where the quantum oscillation and the superconductivity coexist in the presence of the magnetic fields. The examples include the organic superconductor,  $\ensuremath{\kappa}\ensuremath{-}(\mathrm{BEDT}\ensuremath{-}\mathrm{TTF}{)}_{2}\mathrm{Cu}(\mathrm{NCS}{)}_{2}$\cite{PhysRevB.67.144521,PhysRevB.76.024505,PhysRevB.66.224513,PhysRevLett.99.187002,PhysRevLett.88.027002} and high $T_c$ superconductors \cite{Banerjee2013,PhysRevB.95.174503,PhysRevX.9.021025,PhysRevB.97.224520}. In order to understand these experiments, it is important to unveil the generic nature of the superconductivity in the Hofstadter regime.

In the current work, we investigate the universal features of the Hofstadter superconductivity. First of all, to illustrate the general applicability of our work, we demonstrate that the tilted magnetic field applied to a three-dimensional system can generate the Hofstadter butterfly even with the moderate field strength. The van-Hove singularity of the flat bands in the Hofstadter butterfly is shown to enhance the superconducting critical temperature. Furthermore, the superfluid density of the superconductors arising from the Hofstadter butterfly is described by the quantum geometry of the electronic wave function. It turns out that the quantum geometric properties also determine the superconducting nodal structures and the supercurrent. Our discovery of the quantum geometric superconductivity can be applied to generic quasi-two dimensional materials. Finally, based on our theory, we also discuss the case of the field-induced re-entrant superconducting phase in $\textrm{UTe}_2$.

 \begin{figure}
	\includegraphics[scale=0.32]{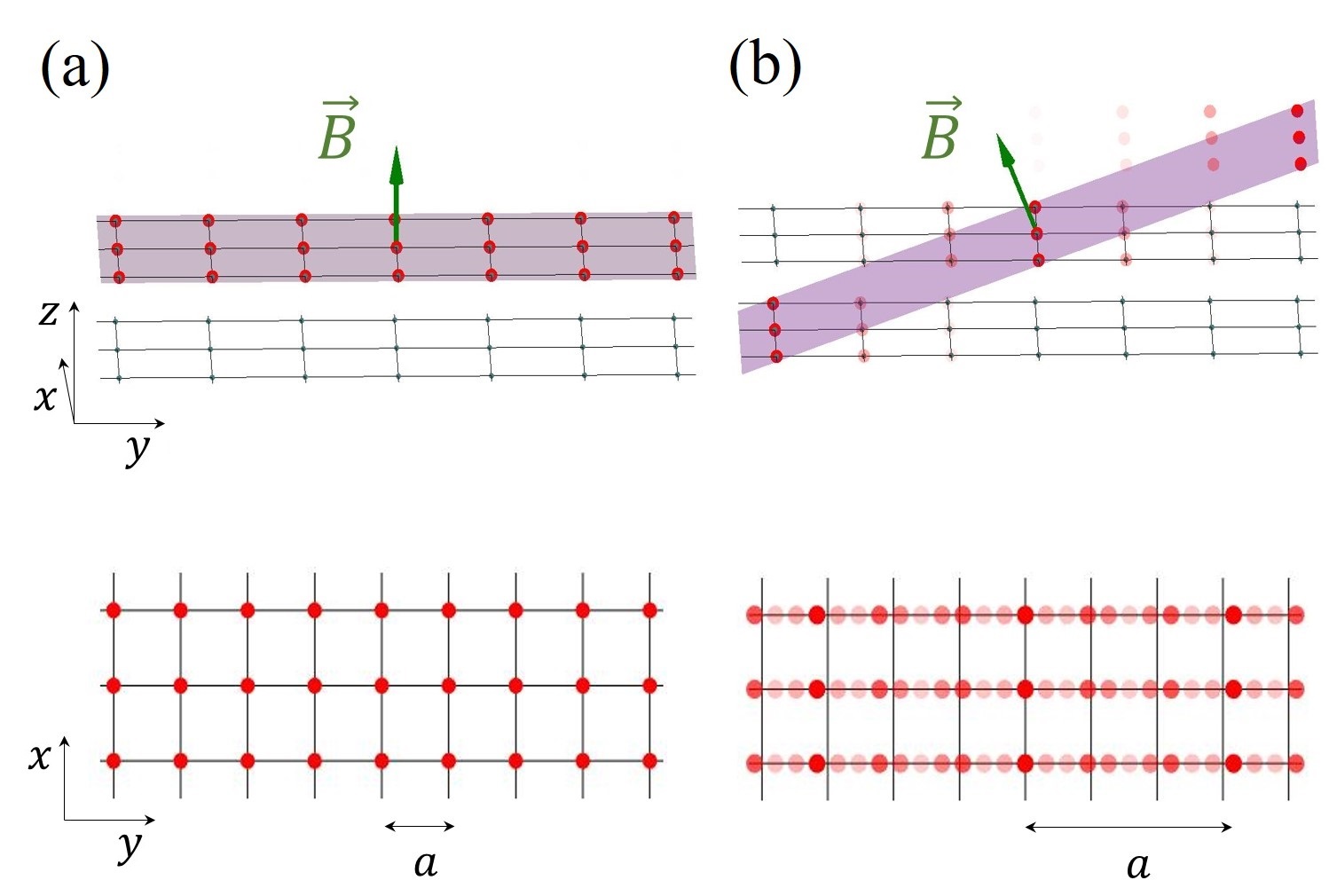}
	\caption{\textbf{Superlattice in the presence of the tilted magnetic field} (a) the field direction is normal to $xy$-plane and (b) the field direction is tilted towards $\hat{y}$-axis. In the tilted field, the projected two-dimensional unit cell forms a superlattice structure (bottom). The superlattice structure can greatly enhance the effective magnetic flux per plaquette, and it realizes the Hofstadter butterfly spectrum with the moderate strength of the magnetic field.}
	\label{Fig:Unit}
\end{figure}

\textbf{3D Butterfly Spectrum}- Our discussion starts with a generic tight-binding model of the quasi-two-dimensional cubic lattice with the nearest neighbor hoppings $T_{x,y,z}$ in each direction($T_{x,y}\gg T_z$). We only assume that the system preserves the time-reversal and the inversion symmetry, which is dictated by the relation, $T_{x,y,z}=T^*_{x,y,z}$. Upon this model, we consider the application of the tilted magnetic field, $\mathbf{B}$, parallel to the $yz$-plane. The corresponding vector potential can be generally written as, $\mathbf{A}=x|\mathbf{B}|(0, \cos\theta,-\sin\theta)$, where $\theta$ is the tilt angle from the $\hat{z}$-axis. Considering the Peierls substitution on the lattice model, we derive the three-dimensional version of the Harper's model with the tilted field\cite{Harper_1955,PhysRevLett.86.1062}, 
\begin{eqnarray}
H(k_y,k_z)_{3D}=\sum_{\langle x,x' \rangle}\textrm{T}_x c^\dagger_{x'} c_x+\sum_{x}\textrm{V}_{0}(k_y,k_z)c^\dagger_{x} c_x,
\label{eq:h0}
\end{eqnarray}
where $\langle,\rangle$ represents the nearest neighbor hopping. The effective potential, $V_0$, is explicitly given as,
\begin{eqnarray}
\textrm{V}_{0}(k_y,k_z)=
2\textrm{T}_y \cos(2\pi \Phi_z x+k_y)+2\textrm{T}_z \cos(2\pi \Phi_y x+k_z ).
\nonumber
\\
\label{Eq:veff}
\end{eqnarray}
where $\Phi_{y(z)}$ is the flux penetrating the $xz \,(xy)$ planes of the unit cell respectively.

If the field direction is normal to the $xy$-plane, the Landau minibands overlap with each other through the $z$-directional dispersion. In this case, the Hofstadter butterfly cannot exist unless the field strength is huge. On the other hand, if the field direction is tilted from the plane, the projected lattice normal to the field direction forms a superlattice structure, which enlarges the effective flux(See Fig. \ref{Fig:Unit}). The effect of the tilted field can be more intuitively understood by analyzing $V_{0}(k_y,k_z)$ of Eq. \eqref{Eq:veff} \cite{PhysRevLett.86.1062,PhysRevB.65.045310}. To do so, we only consider the potential $\textrm{T}_y$ first. In the presence of the flux, $\Phi_z$, the low energy Landau bands form well-localized eigenstates along the $\hat{x}$-direction at the minima of $\textrm{T}_y\cos(2\pi \Phi_z x+k_y)$. The effective $\hat{x}$-directional lattice length is enhanced by $\Phi_z$. Then, we consider the perturbation of the potential, $\textrm{T}_z$. The flux, $\Phi_y$, introduces the additional effective potential to the localized Landau bands, which is given by, $V_{\textrm{eff}}=2T_z \cos(2\pi\Phi_\textrm{eff}n_x-\Phi_\textrm{eff}k_y+k_z)$, where $\Phi_{\textrm{eff}}=\Phi_y/\Phi_z$ is the effective flux, and $n_x$ indicates the enlarged superlattice sites. We find, owing to the superlattice structure, the effective flux is modified and given by the relative ratio between the two fluxes. As a result, the Hofstadter butterfly spectrum can be achieved by tuning the tilt angle $\theta$ even at the moderate strength of the magnetic field (See supplementary material \ref{sec:Spectrum} for the explicit numerical demonstration of the flux magnification with the full tight-binding model.).
 
\begin{figure}
 	\includegraphics[scale=0.55]{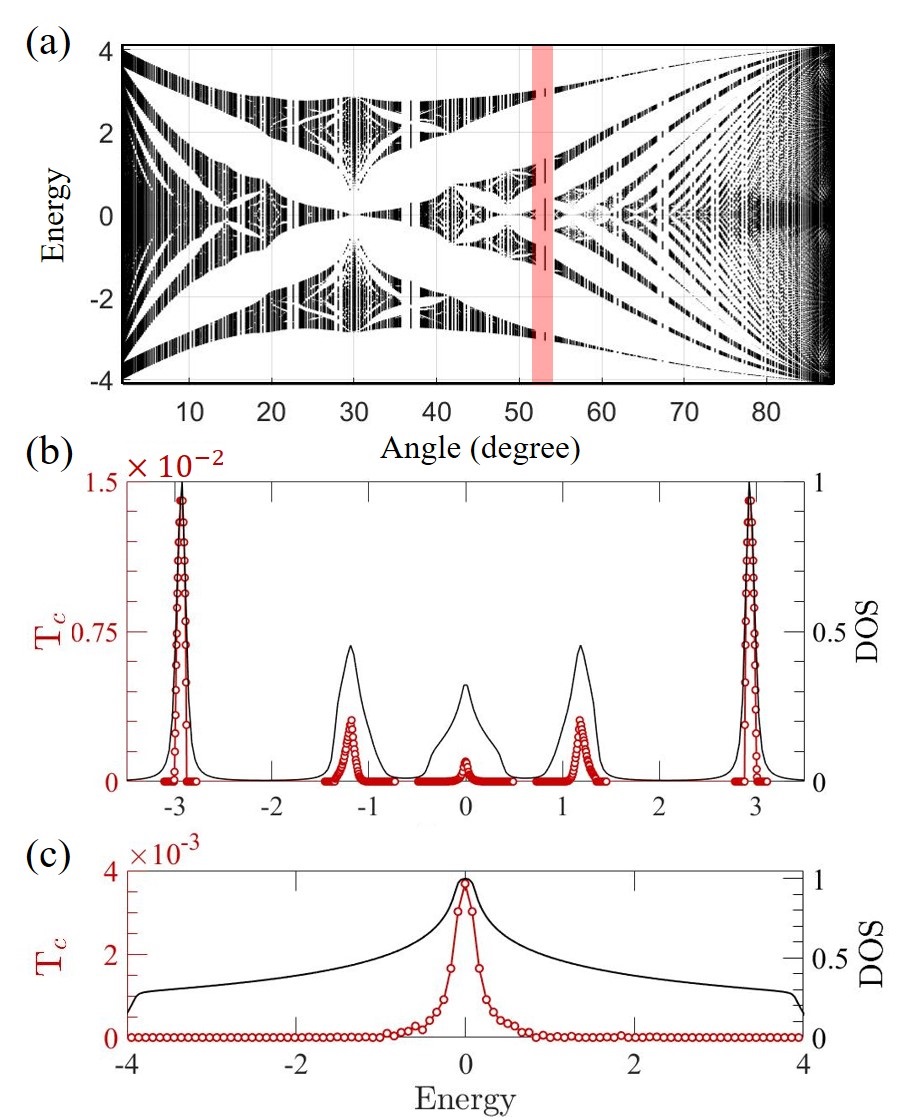}
 	\caption{ \textbf{Enhanced critical temperature in the three dimensional Hofstadter butterfly spectrum} (a) Landau minibands spectrum of the three-dimensional Hofstadter butterfly as a function of the tilt angle($0^\circ$ is normal to $xy$-plane). (b)-(c) The normalized density of states(black line) and $T_c$(red line) as a function of the chemical potential (b) when $(\Phi_{y},\Phi_{z})=(\frac{3}{5},\frac{4}{5})$ and (c) without the field. We find that $T_c$ is enhanced near the vHS of the Landau minibands as compared to the case without the field. We used $\textrm{T}_{x}=\textrm{T}_{y}=1$, $\textrm{T}_{z}=0.05$, and the onsite interaction strength $V=0.6$.}
 	\label{Fig:Tc}
 \end{figure}

\textbf{Field-Enhanced Superconductivity}- After establishing the three-dimensional Hofstadter butterfly system, we now consider the generic form of the pairing interaction in the BCS channel, which can be written as,
\begin{eqnarray}
H_{\textrm{pairing}}=\frac{1}{2}\sum_{\mathbf{k},\mathbf{k}'}V_{\alpha\beta\gamma\delta}(\mathbf{k},\mathbf{k}') c^\dagger_\alpha(\mathbf{k}) c^\dagger_\beta(-\mathbf{k}) c_\gamma(-\mathbf{k}') c_\delta(\mathbf{k}'),
\nonumber
\\
\end{eqnarray}
where the Greek letters represent the spin and the sites in the magnetic unit cell. $V_{\alpha\beta\gamma\delta}(\mathbf{k},\mathbf{k}')$ describes the microscopic interaction. The superconducting critical temperature, $T_c$, can be formally evaluated by solving the self-consistent gap equation\cite{RAJAGOPAL1966539,PhysRevLett.63.2425} with the superconducting pairing kernel, $\mathbf{K}$,
\begin{eqnarray}
\Delta_{\gamma\delta}(\mathbf{k})=\sum_{\mathbf{k}',\lambda\mu}\mathbf{K}_{\gamma\delta,\lambda\mu}(\mathbf{k},\mathbf{k}')\Delta_{\lambda \mu}(\mathbf{k'}),
\label{Eq:selfconsist}
\end{eqnarray}
where the order parameter is defined as, 
$
\Delta_{\gamma\delta}(\mathbf{k})\equiv\sum_{\mathbf{k}',\alpha\beta}V_{\alpha\beta\gamma\delta}(\mathbf{k,k'}) \langle c^\dagger_\alpha(\mathbf{k}') c^\dagger_\beta(-\mathbf{k}') \rangle
\nonumber$. By diagonalizing the full tight-binding model in Eq. \eqref{eq:h0}, one can solve the gap equation. Fig. \ref{Fig:Tc} (b) shows $T_c$ of the singlet pairing with the on-site interaction as a function of the chemical potential. We find that the band flattening greatly enhances $T_c$ compared to the case without a magnetic field. Especially, $T_c$ diverges near the vHSs. We observe that the enhancement of $T_c$ near the vHS of the Landau minibands is a generic feature, independent of the specific form of the interactions. Notice that, in the weak field regime, the Chandrasekhar-Clogston limit\cite{doi:10.1063/1.1777362,PhysRevLett.9.266} would restrain the superconducting $T_c$. Therefore, this field-induced superconductivity would be most likely realized as a re-entrant phase, which arises when the Fermi level crosses the vHSs of the Landau bands. We dub this form of the superconductivity as the three-dimensional Hofstadter butterfly superconductor(3D HBSC). In the following sections, we study the quantum geometrical characters of the 3D HBSC.

\textbf{Geometric Supercurrent}- The superfluid weight, which measures the change of the free energy, $F(\mathbf{q})$, due to the spatially modulating order parameter, $\Delta(\mathbf{r},\mathbf{q})=|\Delta|e^{2i\mathbf{q}\cdot\mathbf{r}}$, characterizes the superfluid behavior of 3D HBSC. The superfluid weight, $\mathbf{D}_{\textrm{s}}$, can be formally evaluated as,
\begin{eqnarray}
[\mathbf{D}_{\textrm{s}}]_{i,j} =\frac{1}{\mathcal{V}}\frac{\partial^2 F(\mathbf{q})}{\partial q_i \partial q_j}
\label{Eq:Ds}
\end{eqnarray}
where $\mathcal{V}$ is the volume of the system and the subscripts $i,j$ indicate the $x,y$ directions. The superfluid weight, $\mathbf{D}_{\textrm{s}}=\mathbf{D}_{\textrm{conv}}+\mathbf{D}_{\textrm{geo}}$, can be decomposed into the two independent contributions: $\mathbf{D}_{\textrm{conv}}\propto \partial_\mathbf{k}^2 \epsilon (\mathbf{k})$, which is proportional to the curvature of the normal bands, $\epsilon(\mathbf{k})$, and $\mathbf{D}_{\textrm{geo}}$, which is the anomalous contribution from the quantum geometry of the wave functions\cite{Peotta2015,PhysRevLett.124.167002,cheng2010quantum,PhysRevB.56.12847,RevModPhys.84.1419,PhysRevLett.117.045303,PhysRevB.95.024515,PhysRevB.98.220504}. 

\begin{figure}
	\includegraphics[scale=0.4]{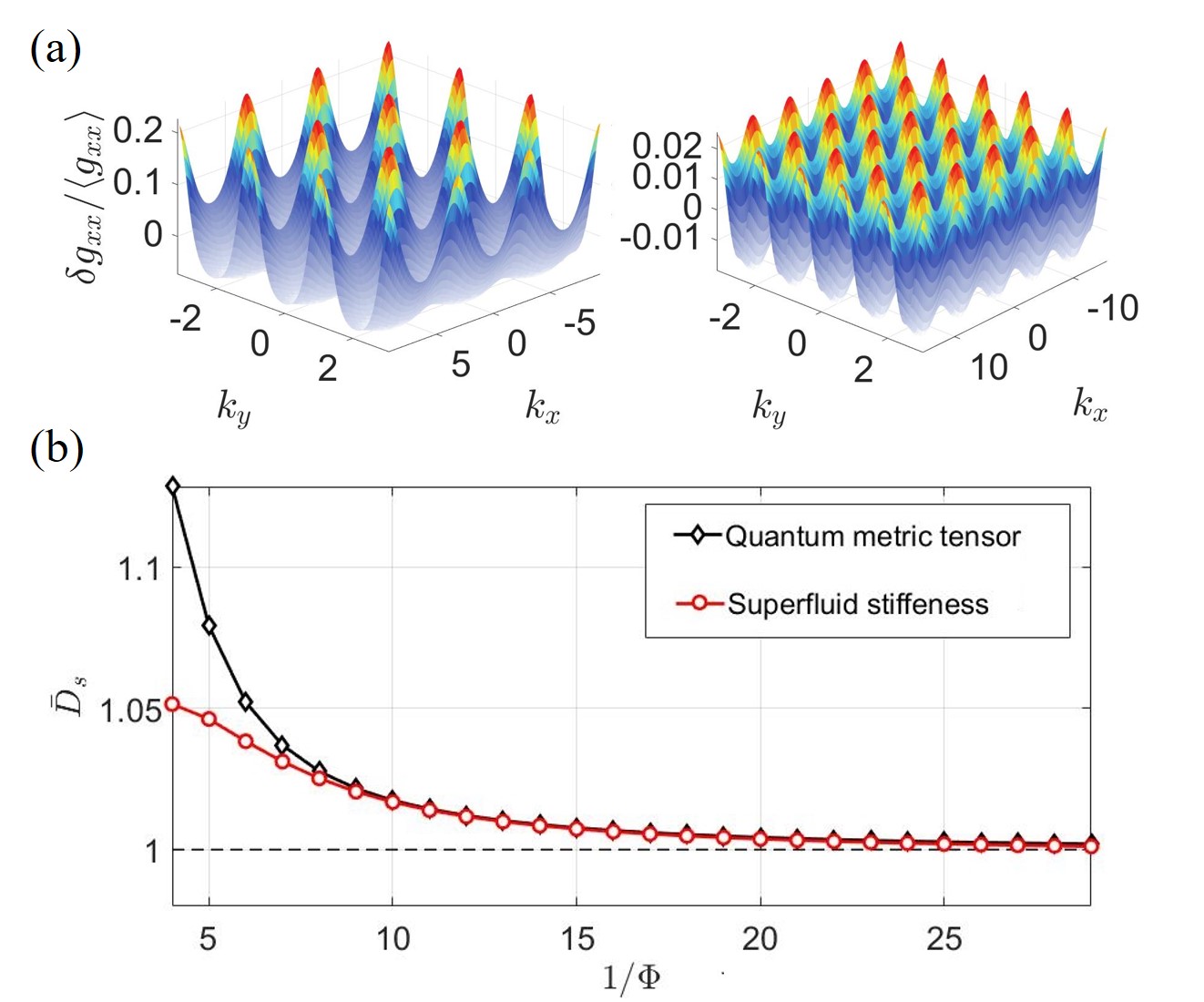}
	\caption{\textbf{Distribution of the quantum geometric tensor and the superfluid weight} (a) the fluctuation about the average of the QGT of the lowest landau band for $\Phi_{\textrm{eff}}=\frac{1}{3}$(left) and $\Phi_{\textrm{eff}}=\frac{1}{5}$(right) respectively. In the low flux limit, the fluctuation decreases, and the distribution asymptotically becomes uniform. (b) Comparison of QGT and the normalized superfluid weight, $\bar{D}_s$,(the normalization is divided by $M(\mk)$ in Eq. \eqref{Eq:Ds2}). In the low flux limit, the contribution of the QGT and the superfluid weight converges due to the band flattening. In this regime, the superfluid weight asymptotically saturates the lower bound.}
	\label{Fig:Ds}
\end{figure}

Furthermore, $\mathbf{D}_{\textrm{geo}}$ is independent of the quasiparticle dispersions, and it is the only contribution to the superfluid weigth if the bands are flat ($\epsilon_\mathbf{k}=\mu$). The situation of the 3D HBSC fits to this case. More precisely, $\mathbf{D}_{\textrm{geo}}$ is dictated by the quantum geometric tensor(QGT)\cite{Peotta2015} (also known as Fubini-Study metric), $\mathbf{g}(\mathbf{k})$, through the following expression,
 
\begin{eqnarray}
[\mathbf{D}_{\textrm{s}}]_{ij}\approx [\mathbf{D}_{\textrm{geo}}]_{ij}= \int_{BZ} \frac{d^3k}{(2\pi)^3}
\textrm{M}(\mathbf{k}) \textrm{Re}[\mathbf{g}(\mathbf{k})]_{ij},
\label{Eq:Ds2}
\end{eqnarray}
where $\textrm{M}(\mathbf{k})=\frac{2|\Delta(\mathbf{k})|^2}{\sqrt{\mu^2+|\Delta(\mathbf{k})|^2}}$ is a form factor. $\Delta(\mathbf{k})$ is the superconducting order parameter of the Landau minibands. The QGT is given in terms of the Bloch wave function $u(\mathbf{k})$ of the Landau miniband as,
\begin{eqnarray}
[\mathbf{g}(\mathbf{k})]_{i,j}&=&\partial_{i} u^\dagger(\mathbf{k})\partial_j u(\mathbf{k})
-\beta_{i}(\mathbf{k})\beta_{j}(\mathbf{k}).
\label{Eq:qgt}
\end{eqnarray}
Here, $\beta_{j}(\mathbf{k}) =iu^\dagger(\mathbf{k})\partial_j u(\mathbf{k})$ corresponds to the Berry connection, and the imaginary component of the QGT corresponds to the Berry curvature\cite{cheng2010quantum}.
(For the detailed derivation of Eq. \eqref{Eq:Ds2}, see the supplementary material \ref{sec:Ds} and \ref{sec:QGT} respectively.) It is particularly important to point out that the positive semi-definiteness of the QGT puts an important lower bound to the diagonal part of the metric such that $[g(\mk)]_{xx,yy} \ge \textrm{Im} [\mathbf{g}(\mathbf{k})_{xy}]$ if $\textrm{T}_x=\textrm{T}_y$ \cite{Peotta2015}. To elaborate this feature, we explicitly calculate the QGT of the lowest Landau minibands.  Fig. \ref{Fig:Ds}(a) shows the momentum distribution of $g_{xx}(\mk)$ as a function of the magnetic flux, $\Phi_{\textrm{eff}}$. In general, the QGT is a function of the momentum. However, we find that the momentum fluctuation of the QGT diminishes as the flux decreases(Fig. \ref{Fig:Ds}(a)). In the low flux limit, the momentum distribution of the QGT becomes uniform. As such, the black line in Fig. \ref{Fig:Ds}(b) shows the asymptotic convergence of the QGT to its theoretical lower bounds, which is nothing more than the Chern number of the Landau minibands. 

Accordingly, the integrand of $\mathbf{D}_{\textrm{s}}$ in Eq. \eqref{Eq:Ds2} reaches to the theoretical lower bound set by the QGT(red line in Fig. \ref{Fig:Ds}(b)). As a result, the superfluid weight also saturates to the lower bound. This result holds regardless of the pairing symmetry of the order parameter since the momentum distribution of the QGT becomes uniform. In conclusion, 3D HBSC realizes the peculiar superfluidity where the supercurrent transport is solely mediated through its quantum geometric channel. 

In contrast to the conventional phenomenology of ordinary superconductors\cite{tinkham2004introduction}, the geometric superfluid weight, $\mathbf{D}_{\textrm{geo}}$, in Eq. \eqref{Eq:Ds2} realizes the nonlinear current-density consecutive relation as,
\begin{eqnarray}
\mathbf{j}_{\textrm{geo}}=\frac{e|\Delta|^2}{\sqrt{\mu^2+|\Delta|^2}}  (\nabla \chi -\frac{2e}{c}\mathbf{A}),
\label{Eq:j}
\end{eqnarray}
where $\chi$ is the phase of the superconducting order parameter. The current density becomes linearly proportional to the order parameter amplitude, as $\mu$ decreases. The nonlinear relation in Eq. \eqref{Eq:j} is the consequence of the flatband superfluidity. We also note that the inclusion of this geometric current, $\mathbf{j}_{\textrm{geo}}$, introduces a finite Meissner effect, which is in sharp contrast to the previous predictions\cite{PhysRevLett.63.2425}. 


\begin{figure}
	\includegraphics[scale=0.5]{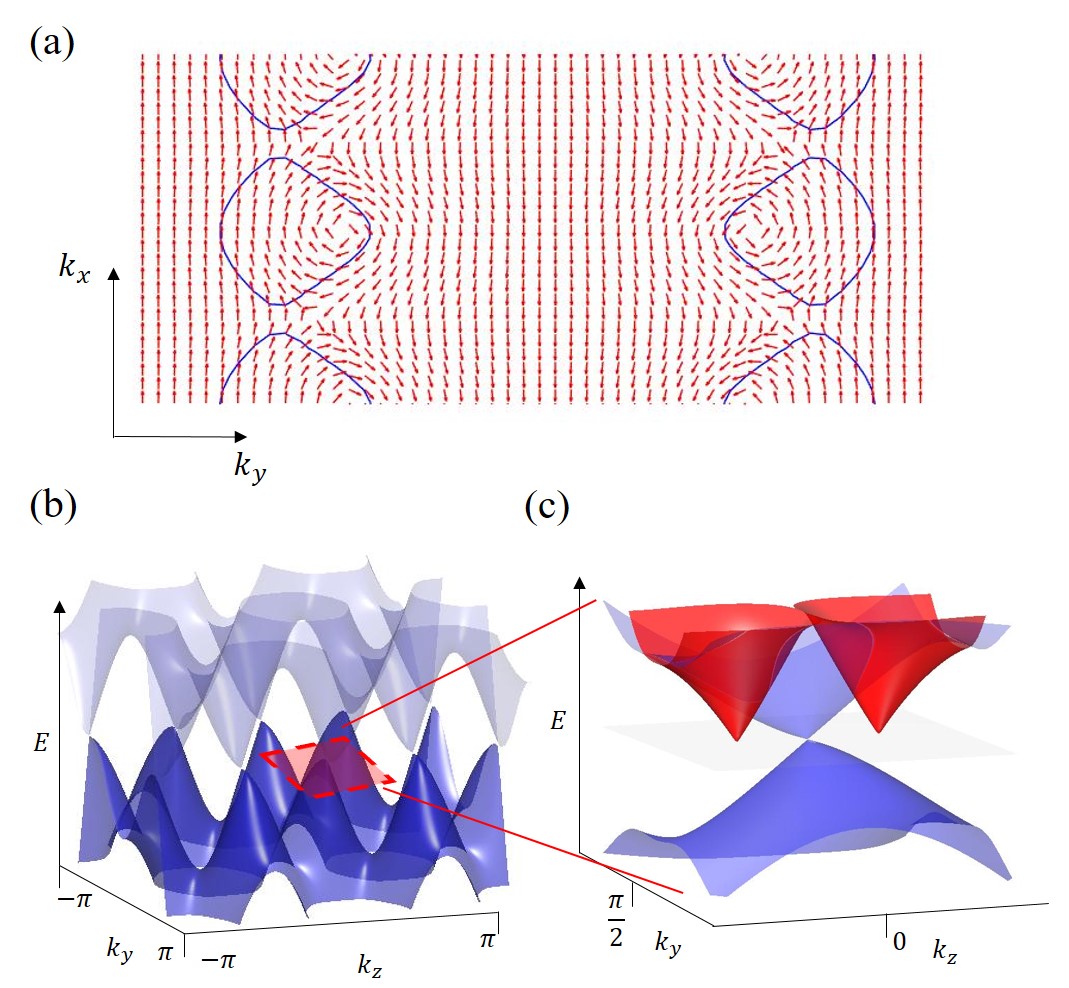}
	\caption{ \textbf{Geometry induced universal superconducting nodal structure}	
	(a) map of the velocity field $\mathbf{v}(\mathbf{k})$ (red arrows) and the Fermi surfaces(blue contours) when $\Phi_y=1/2$, $\Phi_z=1/4$. The QGT ensures the non-zero windings of the velocity field along the Fermi surface.
	(b) Examples of the filling enforced semimetal in 3D Hofstadter butterfly. We find the protected two-fold degeneracy, which forms the Weyl nodes. (c) Superconducting gap structure of (b). The blue surface represents the normal Weyl nodes, which is the monopole of the Berry curvature. The superconducting nodes(tips of the red surface) occur at the location of the vortices of the velocity field on the Fermi surfaces.}
	\label{Fig:Gap}
\end{figure}

\textbf{Universal Nodal Structure}- The quantum geometry of the Landau minibands also plays an important role in the gap structure of the BdG quasiparticle. To see this, we consider the generic form of the pairing potential between the Landau minibands, which is written as $|\Delta(\mathbf{k})|e^{i\phi(\mathbf{k})} a_s(\mathbf{k}) a_{\bar{s}}(\mathbf{-k})+h.c. $, where $a_s(\mathbf{k})$ is the annihilation operator of the Landau minibands near the Fermi level with the spin $s$. We also define the corresponding velocity field of the pairing operator\cite{PhysRevLett.90.057002,PhysRevLett.120.067003}, $\mathbf{v}(\mathbf{k})\equiv \nabla_k \phi(\mathbf{k})-\mathbf{\beta}(\mathbf{k})+\mathbf{\beta}(-\mathbf{k})$. This velocity field is a gauge invariant quantity as it is invariant under the $\textrm{U}(1)$ transformation, $u(\mk)\rightarrow u(\mk)e^{-i\Lambda(\mk)}$, where $\Lambda(\mk)$ is a smooth scalar function. We can consider a winding number along a loop on the two-dimensional normal Fermi surface, which counts the vorticity of the velocity field(See Fig. \ref{Fig:Gap} (a)). The vortical configurations accompany the singularity of the velocity field, which is nothing more than the superconducting nodes($|\Delta(\mathbf{k})|=0$). Interestingly, the QGT of the normal band encapsulates the information of the vorticity of the superconductivity as,
\begin{eqnarray}
\sum_{\textit{C}_i}\oint_{\textit{C}_i} \frac{d\mathbf{k}}{2\pi}\cdot \mathbf{v}(\mathbf{k})
=2\textrm{Im}\int_{\textrm{FS}} dk_i dk_j   [\mathbf{g}(\mathbf{k})-\mathbf{g}(\mathbf{-k})]_{ij},
\nonumber
\\
\label{Eq:g}
\end{eqnarray}
where $\textrm{FS}$ represents the Fermi surface, and $\textrm{C}_i$ represents the one-dimensional loops encircling each vortex. If the Fermi surface possesses a non-zero Chern number, the right hand side of Eq. \eqref{Eq:g} is equal to the difference of the Chern number between the two Fermi surfaces at $\mathbf{k}$ and $\mathbf{-k}$.

For example, we can consider the generic tilt angles with the flux, $\Phi_{y,z}=p_{y,z}/q$, of an even integer denominator, $q\in 2\mathbb{Z}$. The magnetic unit cell contains the even integer sites, and we can define a half-magnetic unit cell translation with the complex conjugation, $\hat{T}_{x,\frac{q}{2}}K$, and the additional momentum translation $k_{y,z}\rightarrow k_{y,z}\pm\pi$, the product of which forms an emergent anti-unitary symmetry of the Hamiltonian in Eq. \eqref{eq:h0}. This symmetry squares to $e^{ik_x}$ due to the half-unit cell translation and gives rise to the additional factor of $-1$ at the $k_x=\pi$ plane. The role of the symmetry is analogous to the nonsymmorphic time-reversal symmetry in the context of the filling enforced semimetal phases\cite{PhysRevLett.118.186401,PhysRevLett.117.096404,PhysRevB.98.184514}. The physical consequence is the modified Kramer's degeneracy at $(\pi,\pm\frac{\pi}{2},\pm\frac{\pi}{2})$\cite{PhysRevLett.118.186401,PhysRevLett.117.096404,PhysRevB.98.184514}. The two-fold point degeneracy in three-dimensions is the Weyl nodes which become the source of the vorticity in Eq. \eqref{Eq:g}. The vortices in normal Fermi surface realize as the Weyl nodes in the BdG spectrum(See Fig. \ref{Fig:Gap} (c)), each of which is protected by the Chern number, $\textrm{C}\in \mathbb{Z}$, in accordance with the class $\textrm{D}$ in the well-known Altland-Zirnbauer classification\cite{Ryu_2010,PhysRevB.82.115120}. As a result, we expect the emergence of the universal Weyl superconductivity controllable by the tilt angle. The Weyl superconductivity is expected to be robust even in the existence of the pair density wave instability and mixed phases\cite{PhysRevB.93.214511,PhysRevB.82.115120,PhysRevLett.121.037701}. 

\textbf{Application to $\textrm{UTe}_2$-}
Finally, we discuss the possible connection of the field polarized re-entrant phase of $\textrm{UTe}_2$ with the 3D HBSC\cite{Ran2019}. $\textrm{UTe}_2$ is the orthorhombic paramagnet, possessing the quasi-two-dimensional Fermi surfaces\cite{PhysRevLett.123.217001,PhysRevLett.124.076401,PhysRevLett.123.217002}. Two re-entrant phases appear near the boundary of the metamagnetic phase transition line. One of the re-entrant phases occurs with the field direction perpendicular to the magnetic easy axis(a-axis), and the other occurs with the tilted field angle $\sim 25^\circ$ on the b-c plane\cite{Ran2019}. In these two re-entrant phases, it is expected that the superconducting order parameters are distinct\cite{knafo2020comparison}, and our theory may explain the latter re-entrant phase with tilted field angle.

Unlike the other uranium-based ferromagnetic superconductors($\textrm{URhGe}$\cite{Aoki2001}, $\textrm{UCoGe}$\cite{PhysRevLett.99.067006}, and $\textrm{UGe}_2$\cite{Saxena2000}), the  field-induced ferromagnetic fluctuation\cite{Mineev_2017} is a less compatible explanation of the re-entrant superconductivity. On the other hand, the metamagnetic transition is shown to accompany several Lifshitz transitions with small hysteresis\cite{PhysRevLett.124.086601}. It may signal the Fermi surface reconstruction associated with the Landau quantization. In such cases, we speculate that the crossing of the vHSs of Landau bands and the Fermi level plays a crucial role in enhancing the superconductivity, thereby realizes as the re-entrant superconductivity at the tilted field. If so, we further speculate that another re-entrant phase dome would appear at higher fields where the next crossing of the Landau level occurs. Moreover, we propose that the effect of the Landau quantization can be experimentally confirmed by the Hall effect measurement near the metamagnetic transition line.

\textbf{Discussions-} We have studied possible superconducting states in a three-dimensional Hofstadter butterfly system. Using the quasi-two-dimensional system subject to the tilted magnetic field, we show that the Hofstadter butterfly spectrum arises from the resulting superlattice structure. It is demonstrated that the superconductivity in the Hofstadter butterfly system possesses a number of intriguing properties that are in stark contrast to the conventional superconductivity. Unlike the superconductivity in the semi-classical regime, the vHSs of the Landau mini-bands in presence of the attractive interaction greatly enhance the superconducting critical temperature. Once the superconductivity is induced in the Hofstadter Landau bands, the superconducting properties are dictated by the quantum geometry of the Landau bands, which is reflected in the geometric supercurrent and universal nodal structure. These behaviors can be experimentally confirmed by the London penetration-depth measurement and thermal Hall measurement\cite{PhysRevB.86.054504,PhysRevLett.121.037701}. We also note that our results are not restricted to the orthorhombic lattice. The tilted field can induce the Hofstadter butterfly, and the enhancement of T$_c$ can also occur in other layered materials such as graphite and bulk transition metal dichalcogenides. Constructing the microscopic theory of the superconductivity in these materials would be an interesting direction for future study.

\section{acknowledgments}
M.J.P. is grateful to Gil Young Cho and GiBaik Sim for fruitful and enlightening discussions.
M.J.P. and S.B.L. are supported by National Research Foundation Grants (NRF-2020R1F1A1073870, NRF-2020R1A4A3079707). Y.B.K. is supported by the NSERC of Canada, the Center for Quantum Materials at the University of Toronto, and the Killam Research Fellowship of the Canada Council for the Arts.

\section*{Competing interests}
The authors declare no conflict of interest.

\section*{Author contributions}
S.B.L. and Y.B.K. conceived and supervised the research. M.J.P. performed the calculations in this work. All authors contributed to writing the manuscript.

\section{Methods}

\subsection{Numerical Calculation of Superconducting Kernel Matrix}
The superconducting critical temperature, $T_c$, can be formally evaluated by solving the self-consistent gap equation\cite{RAJAGOPAL1966539,PhysRevLett.63.2425},
\begin{eqnarray}
\Delta_{\gamma\delta}(\mathbf{k})=\sum_{\mathbf{k}',\lambda\mu}\mathbf{K}_{\gamma\delta,\lambda\mu}(\mathbf{k},\mathbf{k}')\Delta_{\lambda \mu}(\mathbf{k'}),
\label{Eq:selfconsist}
\end{eqnarray}
where the order parameter is defined as, 
$
\Delta_{\gamma\delta}(\mathbf{k})\equiv\sum_{\mathbf{k}',\alpha\beta}V_{\alpha\beta\gamma\delta}(\mathbf{k,k'}) \langle c^\dagger_\alpha(\mathbf{k}') c^\dagger_\beta(-\mathbf{k}') \rangle
\nonumber$.
The superconducting pairing kernel, $\mathbf{K}$, is evaluated as,
\begin{eqnarray}
&\mathbf{K}_{\gamma\delta,\lambda\mu}(\mathbf{k},\mathbf{k}')=
\\
\nonumber
&-T\sum_{\alpha,\beta,k,\omega_n}V_{\alpha\beta\gamma\delta}(\mathbf{k},\mathbf{k}')  [G(\mathbf{k}',\omega_n)]_{\alpha \lambda} [G(-\mathbf{k}',-\omega_n)]_{\beta \mu}.
\end{eqnarray} 
Here $\omega_n$ is the Matsubara frequencies and $G(\mathbf{k},\omega_n)$ is the Matsubara Green function of the normal Hamiltonian in Eq. \eqref{eq:h0}. (See supplementary material \ref{sec:Kernel} for the derivation of $\mathbf{K}$). Eq. \eqref{Eq:selfconsist} takes the form of an eigenvalue problem of the Kernel matrix, where the onset of the superconducting instability occurs if the largest eigenvalue of $\mathbf{K}$ is reached $1$. In the calculation of the critical temperature, we determine $T_c$ by investigating the largest eigenvalue of the Kernel matrix, while gradually decreasing the temperature.

\subsection{Evaluation of Quantum geometric Tensor}

In general, the definition of the QGT in Eq. \eqref{Eq:qgt} is not practical as it requires the gauge fixing to take the derivative of the eigenstates\cite{doi:10.1143/JPSJ.74.1674}. To avoid this, we utilize the following relation,
\begin{eqnarray}
\nonumber
0&=&\partial_\mu (u_{i}^\dagger(\mathbf{k}) H(\mathbf{k}) u_j(\mathbf{k}))
\\ 
\nonumber
&=&
(\epsilon_{j}(\mathbf{k})-\epsilon_{i}(\mathbf{k}))(\partial_\mu u_{i}^\dagger(\mathbf{k}) )u_j(\mathbf{k})
+u_{i}^\dagger(\mathbf{k}) \partial_{\mu} H(\mathbf{k}) u_j(\mathbf{k}),
\\
&&\partial_\mu u_{i}^\dagger(\mathbf{k}) u_j(\mathbf{k})
= \frac{u_{i}^\dagger(\mathbf{k}) \partial_{\mu} H(\mathbf{k}) u_j(\mathbf{k})}{(\epsilon_{i}(\mathbf{k})-\epsilon_{j}(\mathbf{k}))}.
\end{eqnarray}
The QGT in Eq. \eqref{Eq:qgt} can be rewritten as, 
\begin{eqnarray}
g_{\mu\nu}(\mathbf{k})=\sum_{j\neq i}
\frac{[u_{i}^\dagger(\mathbf{k}) \partial_{\mu} H(\mathbf{k}) u_j(\mathbf{k})][u_{j}^\dagger(\mathbf{k}) \partial_{\mu} H(\mathbf{k}) u_i(\mathbf{k})]}{(\epsilon_{i}(\mathbf{k})-\epsilon_{j}(\mathbf{k}))^2}
\nonumber
\\
\label{eq:Sgdef2}
\end{eqnarray}
Now the derivative of the Hamiltonian does not have gauge ambiguity. We have utilized Eq. \eqref{eq:Sgdef2} to numerically evaluate the QGT of the Landa minibands.

\section{Data availability}
The data that support the findings of this study are available from the corresponding author upon request.

\bibliography{reference}

\begin{thebibliography}{63}%
\makeatletter
\providecommand \@ifxundefined [1]{%
 \@ifx{#1\undefined}
}%
\providecommand \@ifnum [1]{%
 \ifnum #1\expandafter \@firstoftwo
 \else \expandafter \@secondoftwo
 \fi
}%
\providecommand \@ifx [1]{%
 \ifx #1\expandafter \@firstoftwo
 \else \expandafter \@secondoftwo
 \fi
}%
\providecommand \natexlab [1]{#1}%
\providecommand \enquote  [1]{``#1''}%
\providecommand \bibnamefont  [1]{#1}%
\providecommand \bibfnamefont [1]{#1}%
\providecommand \citenamefont [1]{#1}%
\providecommand \href@noop [0]{\@secondoftwo}%
\providecommand \href [0]{\begingroup \@sanitize@url \@href}%
\providecommand \@href[1]{\@@startlink{#1}\@@href}%
\providecommand \@@href[1]{\endgroup#1\@@endlink}%
\providecommand \@sanitize@url [0]{\catcode `\\12\catcode `\$12\catcode
  `\&12\catcode `\#12\catcode `\^12\catcode `\_12\catcode `\%12\relax}%
\providecommand \@@startlink[1]{}%
\providecommand \@@endlink[0]{}%
\providecommand \url  [0]{\begingroup\@sanitize@url \@url }%
\providecommand \@url [1]{\endgroup\@href {#1}{\urlprefix }}%
\providecommand \urlprefix  [0]{URL }%
\providecommand \Eprint [0]{\href }%
\providecommand \doibase [0]{http://dx.doi.org/}%
\providecommand \selectlanguage [0]{\@gobble}%
\providecommand \bibinfo  [0]{\@secondoftwo}%
\providecommand \bibfield  [0]{\@secondoftwo}%
\providecommand \translation [1]{[#1]}%
\providecommand \BibitemOpen [0]{}%
\providecommand \bibitemStop [0]{}%
\providecommand \bibitemNoStop [0]{.\EOS\space}%
\providecommand \EOS [0]{\spacefactor3000\relax}%
\providecommand \BibitemShut  [1]{\csname bibitem#1\endcsname}%
\let\auto@bib@innerbib\@empty
\bibitem [{\citenamefont {Gor’kov}(1959)}]{gor1959microscopic}%
  \BibitemOpen
  \bibfield  {author} {\bibinfo {author} {\bibfnamefont {Lev~Petrovich}\
  \bibnamefont {Gor’kov}},\ }\bibfield  {title} {\enquote {\bibinfo {title}
  {Microscopic derivation of the ginzburg-landau equations in the theory of
  superconductivity},}\ }\href@noop {} {\bibfield  {journal} {\bibinfo
  {journal} {Sov. Phys. JETP}\ }\textbf {\bibinfo {volume} {9}},\ \bibinfo
  {pages} {1364--1367} (\bibinfo {year} {1959})}\BibitemShut {NoStop}%
\bibitem [{\citenamefont {Eilenberger}(1968)}]{eilenberger1968transformation}%
  \BibitemOpen
  \bibfield  {author} {\bibinfo {author} {\bibfnamefont {Gert}\ \bibnamefont
  {Eilenberger}},\ }\bibfield  {title} {\enquote {\bibinfo {title}
  {Transformation of gorkov's equation for type ii superconductors into
  transport-like equations},}\ }\href@noop {} {\bibfield  {journal} {\bibinfo
  {journal} {Zeitschrift f{\"u}r Physik A Hadrons and nuclei}\ }\textbf
  {\bibinfo {volume} {214}},\ \bibinfo {pages} {195--213} (\bibinfo {year}
  {1968})}\BibitemShut {NoStop}%
\bibitem [{\citenamefont {Hirsch}\ and\ \citenamefont
  {Scalapino}(1986)}]{PhysRevLett.56.2732}%
  \BibitemOpen
  \bibfield  {author} {\bibinfo {author} {\bibfnamefont {J.~E.}\ \bibnamefont
  {Hirsch}}\ and\ \bibinfo {author} {\bibfnamefont {D.~J.}\ \bibnamefont
  {Scalapino}},\ }\bibfield  {title} {\enquote {\bibinfo {title} {Enhanced
  superconductivity in quasi two-dimensional systems},}\ }\href {\doibase
  10.1103/PhysRevLett.56.2732} {\bibfield  {journal} {\bibinfo  {journal}
  {Phys. Rev. Lett.}\ }\textbf {\bibinfo {volume} {56}},\ \bibinfo {pages}
  {2732--2735} (\bibinfo {year} {1986})}\BibitemShut {NoStop}%
\bibitem [{\citenamefont {Markiewicz}(1997)}]{MARKIEWICZ19971179}%
  \BibitemOpen
  \bibfield  {author} {\bibinfo {author} {\bibfnamefont {R.S}\ \bibnamefont
  {Markiewicz}},\ }\bibfield  {title} {\enquote {\bibinfo {title} {A survey of
  the van hove scenario for high-tc superconductivity with special emphasis on
  pseudogaps and striped phases},}\ }\href {\doibase
  https://doi.org/10.1016/S0022-3697(97)00025-5} {\bibfield  {journal}
  {\bibinfo  {journal} {Journal of Physics and Chemistry of Solids}\ }\textbf
  {\bibinfo {volume} {58}},\ \bibinfo {pages} {1179 -- 1310} (\bibinfo {year}
  {1997})}\BibitemShut {NoStop}%
\bibitem [{\citenamefont {Song}\ and\ \citenamefont
  {Koshelev}(2019)}]{PhysRevX.9.021025}%
  \BibitemOpen
  \bibfield  {author} {\bibinfo {author} {\bibfnamefont {Kok~Wee}\ \bibnamefont
  {Song}}\ and\ \bibinfo {author} {\bibfnamefont {Alexei~E.}\ \bibnamefont
  {Koshelev}},\ }\bibfield  {title} {\enquote {\bibinfo {title} {Quantum fflo
  state in clean layered superconductors},}\ }\href {\doibase
  10.1103/PhysRevX.9.021025} {\bibfield  {journal} {\bibinfo  {journal} {Phys.
  Rev. X}\ }\textbf {\bibinfo {volume} {9}},\ \bibinfo {pages} {021025}
  (\bibinfo {year} {2019})}\BibitemShut {NoStop}%
\bibitem [{\citenamefont {Gruenberg}\ and\ \citenamefont
  {Gunther}(1966)}]{PhysRevLett.16.996}%
  \BibitemOpen
  \bibfield  {author} {\bibinfo {author} {\bibfnamefont {Leonard~W.}\
  \bibnamefont {Gruenberg}}\ and\ \bibinfo {author} {\bibfnamefont {Leon}\
  \bibnamefont {Gunther}},\ }\bibfield  {title} {\enquote {\bibinfo {title}
  {Fulde-ferrell effect in type-ii superconductors},}\ }\href {\doibase
  10.1103/PhysRevLett.16.996} {\bibfield  {journal} {\bibinfo  {journal} {Phys.
  Rev. Lett.}\ }\textbf {\bibinfo {volume} {16}},\ \bibinfo {pages} {996--998}
  (\bibinfo {year} {1966})}\BibitemShut {NoStop}%
\bibitem [{\citenamefont {Hofstadter}(1976)}]{PhysRevB.14.2239}%
  \BibitemOpen
  \bibfield  {author} {\bibinfo {author} {\bibfnamefont {Douglas~R.}\
  \bibnamefont {Hofstadter}},\ }\bibfield  {title} {\enquote {\bibinfo {title}
  {Energy levels and wave functions of bloch electrons in rational and
  irrational magnetic fields},}\ }\href {\doibase 10.1103/PhysRevB.14.2239}
  {\bibfield  {journal} {\bibinfo  {journal} {Phys. Rev. B}\ }\textbf {\bibinfo
  {volume} {14}},\ \bibinfo {pages} {2239--2249} (\bibinfo {year}
  {1976})}\BibitemShut {NoStop}%
\bibitem [{\citenamefont {Ran}\ \emph {et~al.}(2019{\natexlab{a}})\citenamefont
  {Ran}, \citenamefont {Eckberg}, \citenamefont {Ding}, \citenamefont
  {Furukawa}, \citenamefont {Metz}, \citenamefont {Saha}, \citenamefont {Liu},
  \citenamefont {Zic}, \citenamefont {Kim}, \citenamefont {Paglione},\ and\
  \citenamefont {Butch}}]{Ran684}%
  \BibitemOpen
  \bibfield  {author} {\bibinfo {author} {\bibfnamefont {Sheng}\ \bibnamefont
  {Ran}}, \bibinfo {author} {\bibfnamefont {Chris}\ \bibnamefont {Eckberg}},
  \bibinfo {author} {\bibfnamefont {Qing-Ping}\ \bibnamefont {Ding}}, \bibinfo
  {author} {\bibfnamefont {Yuji}\ \bibnamefont {Furukawa}}, \bibinfo {author}
  {\bibfnamefont {Tristin}\ \bibnamefont {Metz}}, \bibinfo {author}
  {\bibfnamefont {Shanta~R.}\ \bibnamefont {Saha}}, \bibinfo {author}
  {\bibfnamefont {I-Lin}\ \bibnamefont {Liu}}, \bibinfo {author} {\bibfnamefont
  {Mark}\ \bibnamefont {Zic}}, \bibinfo {author} {\bibfnamefont {Hyunsoo}\
  \bibnamefont {Kim}}, \bibinfo {author} {\bibfnamefont {Johnpierre}\
  \bibnamefont {Paglione}}, \ and\ \bibinfo {author} {\bibfnamefont
  {Nicholas~P.}\ \bibnamefont {Butch}},\ }\bibfield  {title} {\enquote
  {\bibinfo {title} {Nearly ferromagnetic spin-triplet superconductivity},}\
  }\href {\doibase 10.1126/science.aav8645} {\bibfield  {journal} {\bibinfo
  {journal} {Science}\ }\textbf {\bibinfo {volume} {365}},\ \bibinfo {pages}
  {684--687} (\bibinfo {year} {2019}{\natexlab{a}})},\ \Eprint
  {http://arxiv.org/abs/https://science.sciencemag.org/content/365/6454/684.full.pdf}
  {https://science.sciencemag.org/content/365/6454/684.full.pdf} \BibitemShut
  {NoStop}%
\bibitem [{\citenamefont {Aoki}\ \emph {et~al.}(2019)\citenamefont {Aoki},
  \citenamefont {Nakamura}, \citenamefont {Honda}, \citenamefont {Li},
  \citenamefont {Homma}, \citenamefont {Shimizu}, \citenamefont {Sato},
  \citenamefont {Knebel}, \citenamefont {Brison}, \citenamefont {Pourret},
  \citenamefont {Braithwaite}, \citenamefont {Lapertot}, \citenamefont {Niu},
  \citenamefont {Vališka}, \citenamefont {Harima},\ and\ \citenamefont
  {Flouquet}}]{doi:10.7566/JPSJ.88.043702}%
  \BibitemOpen
  \bibfield  {author} {\bibinfo {author} {\bibfnamefont {Dai}\ \bibnamefont
  {Aoki}}, \bibinfo {author} {\bibfnamefont {Ai}~\bibnamefont {Nakamura}},
  \bibinfo {author} {\bibfnamefont {Fuminori}\ \bibnamefont {Honda}}, \bibinfo
  {author} {\bibfnamefont {DeXin}\ \bibnamefont {Li}}, \bibinfo {author}
  {\bibfnamefont {Yoshiya}\ \bibnamefont {Homma}}, \bibinfo {author}
  {\bibfnamefont {Yusei}\ \bibnamefont {Shimizu}}, \bibinfo {author}
  {\bibfnamefont {Yoshiki~J.}\ \bibnamefont {Sato}}, \bibinfo {author}
  {\bibfnamefont {Georg}\ \bibnamefont {Knebel}}, \bibinfo {author}
  {\bibfnamefont {Jean-Pascal}\ \bibnamefont {Brison}}, \bibinfo {author}
  {\bibfnamefont {Alexandre}\ \bibnamefont {Pourret}}, \bibinfo {author}
  {\bibfnamefont {Daniel}\ \bibnamefont {Braithwaite}}, \bibinfo {author}
  {\bibfnamefont {Gerard}\ \bibnamefont {Lapertot}}, \bibinfo {author}
  {\bibfnamefont {Qun}\ \bibnamefont {Niu}}, \bibinfo {author} {\bibfnamefont
  {Michal}\ \bibnamefont {Vališka}}, \bibinfo {author} {\bibfnamefont
  {Hisatomo}\ \bibnamefont {Harima}}, \ and\ \bibinfo {author} {\bibfnamefont
  {Jacques}\ \bibnamefont {Flouquet}},\ }\bibfield  {title} {\enquote {\bibinfo
  {title} {Unconventional superconductivity in heavy fermion ute2},}\ }\href
  {\doibase 10.7566/JPSJ.88.043702} {\bibfield  {journal} {\bibinfo  {journal}
  {Journal of the Physical Society of Japan}\ }\textbf {\bibinfo {volume}
  {88}},\ \bibinfo {pages} {043702} (\bibinfo {year} {2019})},\ \Eprint
  {http://arxiv.org/abs/https://doi.org/10.7566/JPSJ.88.043702}
  {https://doi.org/10.7566/JPSJ.88.043702} \BibitemShut {NoStop}%
\bibitem [{\citenamefont {Miyake}\ \emph {et~al.}(2019)\citenamefont {Miyake},
  \citenamefont {Shimizu}, \citenamefont {Sato}, \citenamefont {Li},
  \citenamefont {Nakamura}, \citenamefont {Homma}, \citenamefont {Honda},
  \citenamefont {Flouquet}, \citenamefont {Tokunaga},\ and\ \citenamefont
  {Aoki}}]{doi:10.7566/JPSJ.88.063706}%
  \BibitemOpen
  \bibfield  {author} {\bibinfo {author} {\bibfnamefont {Atsushi}\ \bibnamefont
  {Miyake}}, \bibinfo {author} {\bibfnamefont {Yusei}\ \bibnamefont {Shimizu}},
  \bibinfo {author} {\bibfnamefont {Yoshiki~J.}\ \bibnamefont {Sato}}, \bibinfo
  {author} {\bibfnamefont {Dexin}\ \bibnamefont {Li}}, \bibinfo {author}
  {\bibfnamefont {Ai}~\bibnamefont {Nakamura}}, \bibinfo {author}
  {\bibfnamefont {Yoshiya}\ \bibnamefont {Homma}}, \bibinfo {author}
  {\bibfnamefont {Fuminori}\ \bibnamefont {Honda}}, \bibinfo {author}
  {\bibfnamefont {Jacques}\ \bibnamefont {Flouquet}}, \bibinfo {author}
  {\bibfnamefont {Masashi}\ \bibnamefont {Tokunaga}}, \ and\ \bibinfo {author}
  {\bibfnamefont {Dai}\ \bibnamefont {Aoki}},\ }\bibfield  {title} {\enquote
  {\bibinfo {title} {Metamagnetic transition in heavy fermion superconductor
  ute2},}\ }\href {\doibase 10.7566/JPSJ.88.063706} {\bibfield  {journal}
  {\bibinfo  {journal} {Journal of the Physical Society of Japan}\ }\textbf
  {\bibinfo {volume} {88}},\ \bibinfo {pages} {063706} (\bibinfo {year}
  {2019})},\ \Eprint
  {http://arxiv.org/abs/https://doi.org/10.7566/JPSJ.88.063706}
  {https://doi.org/10.7566/JPSJ.88.063706} \BibitemShut {NoStop}%
\bibitem [{\citenamefont {Ran}\ \emph {et~al.}(2019{\natexlab{b}})\citenamefont
  {Ran}, \citenamefont {Liu}, \citenamefont {Eo}, \citenamefont {Campbell},
  \citenamefont {Neves}, \citenamefont {Fuhrman}, \citenamefont {Saha},
  \citenamefont {Eckberg}, \citenamefont {Kim}, \citenamefont {Graf},
  \citenamefont {Balakirev}, \citenamefont {Singleton}, \citenamefont
  {Paglione},\ and\ \citenamefont {Butch}}]{Ran2019}%
  \BibitemOpen
  \bibfield  {author} {\bibinfo {author} {\bibfnamefont {Sheng}\ \bibnamefont
  {Ran}}, \bibinfo {author} {\bibfnamefont {I-Lin}\ \bibnamefont {Liu}},
  \bibinfo {author} {\bibfnamefont {Yun~Suk}\ \bibnamefont {Eo}}, \bibinfo
  {author} {\bibfnamefont {Daniel~J.}\ \bibnamefont {Campbell}}, \bibinfo
  {author} {\bibfnamefont {Paul~M.}\ \bibnamefont {Neves}}, \bibinfo {author}
  {\bibfnamefont {Wesley~T.}\ \bibnamefont {Fuhrman}}, \bibinfo {author}
  {\bibfnamefont {Shanta~R.}\ \bibnamefont {Saha}}, \bibinfo {author}
  {\bibfnamefont {Christopher}\ \bibnamefont {Eckberg}}, \bibinfo {author}
  {\bibfnamefont {Hyunsoo}\ \bibnamefont {Kim}}, \bibinfo {author}
  {\bibfnamefont {David}\ \bibnamefont {Graf}}, \bibinfo {author}
  {\bibfnamefont {Fedor}\ \bibnamefont {Balakirev}}, \bibinfo {author}
  {\bibfnamefont {John}\ \bibnamefont {Singleton}}, \bibinfo {author}
  {\bibfnamefont {Johnpierre}\ \bibnamefont {Paglione}}, \ and\ \bibinfo
  {author} {\bibfnamefont {Nicholas~P.}\ \bibnamefont {Butch}},\ }\bibfield
  {title} {\enquote {\bibinfo {title} {Extreme magnetic field-boosted
  superconductivity},}\ }\href {\doibase 10.1038/s41567-019-0670-x} {\bibfield
  {journal} {\bibinfo  {journal} {Nature Physics}\ }\textbf {\bibinfo {volume}
  {15}},\ \bibinfo {pages} {1250--1254} (\bibinfo {year}
  {2019}{\natexlab{b}})}\BibitemShut {NoStop}%
\bibitem [{\citenamefont {Sundar}\ \emph {et~al.}(2019)\citenamefont {Sundar},
  \citenamefont {Gheidi}, \citenamefont {Akintola}, \citenamefont {C\^ot\'e},
  \citenamefont {Dunsiger}, \citenamefont {Ran}, \citenamefont {Butch},
  \citenamefont {Saha}, \citenamefont {Paglione},\ and\ \citenamefont
  {Sonier}}]{PhysRevB.100.140502}%
  \BibitemOpen
  \bibfield  {author} {\bibinfo {author} {\bibfnamefont {Shyam}\ \bibnamefont
  {Sundar}}, \bibinfo {author} {\bibfnamefont {S.}~\bibnamefont {Gheidi}},
  \bibinfo {author} {\bibfnamefont {K.}~\bibnamefont {Akintola}}, \bibinfo
  {author} {\bibfnamefont {A.~M.}\ \bibnamefont {C\^ot\'e}}, \bibinfo {author}
  {\bibfnamefont {S.~R.}\ \bibnamefont {Dunsiger}}, \bibinfo {author}
  {\bibfnamefont {S.}~\bibnamefont {Ran}}, \bibinfo {author} {\bibfnamefont
  {N.~P.}\ \bibnamefont {Butch}}, \bibinfo {author} {\bibfnamefont {S.~R.}\
  \bibnamefont {Saha}}, \bibinfo {author} {\bibfnamefont {J.}~\bibnamefont
  {Paglione}}, \ and\ \bibinfo {author} {\bibfnamefont {J.~E.}\ \bibnamefont
  {Sonier}},\ }\bibfield  {title} {\enquote {\bibinfo {title} {Coexistence of
  ferromagnetic fluctuations and superconductivity in the actinide
  superconductor ${\mathrm{ute}}_{2}$},}\ }\href {\doibase
  10.1103/PhysRevB.100.140502} {\bibfield  {journal} {\bibinfo  {journal}
  {Phys. Rev. B}\ }\textbf {\bibinfo {volume} {100}},\ \bibinfo {pages}
  {140502} (\bibinfo {year} {2019})}\BibitemShut {NoStop}%
\bibitem [{\citenamefont {Tokunaga}\ \emph {et~al.}(2019)\citenamefont
  {Tokunaga}, \citenamefont {Sakai}, \citenamefont {Kambe}, \citenamefont
  {Hattori}, \citenamefont {Higa}, \citenamefont {Nakamine}, \citenamefont
  {Kitagawa}, \citenamefont {Ishida}, \citenamefont {Nakamura}, \citenamefont
  {Shimizu}, \citenamefont {Homma}, \citenamefont {Li}, \citenamefont {Honda},\
  and\ \citenamefont {Aoki}}]{doi:10.7566/JPSJ.88.073701}%
  \BibitemOpen
  \bibfield  {author} {\bibinfo {author} {\bibfnamefont {Yo}~\bibnamefont
  {Tokunaga}}, \bibinfo {author} {\bibfnamefont {Hironori}\ \bibnamefont
  {Sakai}}, \bibinfo {author} {\bibfnamefont {Shinsaku}\ \bibnamefont {Kambe}},
  \bibinfo {author} {\bibfnamefont {Taisuke}\ \bibnamefont {Hattori}}, \bibinfo
  {author} {\bibfnamefont {Nonoka}\ \bibnamefont {Higa}}, \bibinfo {author}
  {\bibfnamefont {Genki}\ \bibnamefont {Nakamine}}, \bibinfo {author}
  {\bibfnamefont {Shunsaku}\ \bibnamefont {Kitagawa}}, \bibinfo {author}
  {\bibfnamefont {Kenji}\ \bibnamefont {Ishida}}, \bibinfo {author}
  {\bibfnamefont {Ai}~\bibnamefont {Nakamura}}, \bibinfo {author}
  {\bibfnamefont {Yusei}\ \bibnamefont {Shimizu}}, \bibinfo {author}
  {\bibfnamefont {Yoshiya}\ \bibnamefont {Homma}}, \bibinfo {author}
  {\bibfnamefont {DeXin}\ \bibnamefont {Li}}, \bibinfo {author} {\bibfnamefont
  {Fuminori}\ \bibnamefont {Honda}}, \ and\ \bibinfo {author} {\bibfnamefont
  {Dai}\ \bibnamefont {Aoki}},\ }\bibfield  {title} {\enquote {\bibinfo {title}
  {125te-nmr study on a single crystal of heavy fermion superconductor ute2},}\
  }\href {\doibase 10.7566/JPSJ.88.073701} {\bibfield  {journal} {\bibinfo
  {journal} {Journal of the Physical Society of Japan}\ }\textbf {\bibinfo
  {volume} {88}},\ \bibinfo {pages} {073701} (\bibinfo {year} {2019})},\
  \Eprint {http://arxiv.org/abs/https://doi.org/10.7566/JPSJ.88.073701}
  {https://doi.org/10.7566/JPSJ.88.073701} \BibitemShut {NoStop}%
\bibitem [{\citenamefont {Niu}\ \emph {et~al.}(2020)\citenamefont {Niu},
  \citenamefont {Knebel}, \citenamefont {Braithwaite}, \citenamefont {Aoki},
  \citenamefont {Lapertot}, \citenamefont {Seyfarth}, \citenamefont {Brison},
  \citenamefont {Flouquet},\ and\ \citenamefont
  {Pourret}}]{PhysRevLett.124.086601}%
  \BibitemOpen
  \bibfield  {author} {\bibinfo {author} {\bibfnamefont {Q.}~\bibnamefont
  {Niu}}, \bibinfo {author} {\bibfnamefont {G.}~\bibnamefont {Knebel}},
  \bibinfo {author} {\bibfnamefont {D.}~\bibnamefont {Braithwaite}}, \bibinfo
  {author} {\bibfnamefont {D.}~\bibnamefont {Aoki}}, \bibinfo {author}
  {\bibfnamefont {G.}~\bibnamefont {Lapertot}}, \bibinfo {author}
  {\bibfnamefont {G.}~\bibnamefont {Seyfarth}}, \bibinfo {author}
  {\bibfnamefont {J-P.}\ \bibnamefont {Brison}}, \bibinfo {author}
  {\bibfnamefont {J.}~\bibnamefont {Flouquet}}, \ and\ \bibinfo {author}
  {\bibfnamefont {A.}~\bibnamefont {Pourret}},\ }\bibfield  {title} {\enquote
  {\bibinfo {title} {Fermi-surface instability in the heavy-fermion
  superconductor ${\mathrm{ute}}_{2}$},}\ }\href {\doibase
  10.1103/PhysRevLett.124.086601} {\bibfield  {journal} {\bibinfo  {journal}
  {Phys. Rev. Lett.}\ }\textbf {\bibinfo {volume} {124}},\ \bibinfo {pages}
  {086601} (\bibinfo {year} {2020})}\BibitemShut {NoStop}%
\bibitem [{\citenamefont {Metz}\ \emph {et~al.}(2019)\citenamefont {Metz},
  \citenamefont {Bae}, \citenamefont {Ran}, \citenamefont {Liu}, \citenamefont
  {Eo}, \citenamefont {Fuhrman}, \citenamefont {Agterberg}, \citenamefont
  {Anlage}, \citenamefont {Butch},\ and\ \citenamefont
  {Paglione}}]{PhysRevB.100.220504}%
  \BibitemOpen
  \bibfield  {author} {\bibinfo {author} {\bibfnamefont {Tristin}\ \bibnamefont
  {Metz}}, \bibinfo {author} {\bibfnamefont {Seokjin}\ \bibnamefont {Bae}},
  \bibinfo {author} {\bibfnamefont {Sheng}\ \bibnamefont {Ran}}, \bibinfo
  {author} {\bibfnamefont {I-Lin}\ \bibnamefont {Liu}}, \bibinfo {author}
  {\bibfnamefont {Yun~Suk}\ \bibnamefont {Eo}}, \bibinfo {author}
  {\bibfnamefont {Wesley~T.}\ \bibnamefont {Fuhrman}}, \bibinfo {author}
  {\bibfnamefont {Daniel~F.}\ \bibnamefont {Agterberg}}, \bibinfo {author}
  {\bibfnamefont {Steven~M.}\ \bibnamefont {Anlage}}, \bibinfo {author}
  {\bibfnamefont {Nicholas~P.}\ \bibnamefont {Butch}}, \ and\ \bibinfo {author}
  {\bibfnamefont {Johnpierre}\ \bibnamefont {Paglione}},\ }\bibfield  {title}
  {\enquote {\bibinfo {title} {Point-node gap structure of the spin-triplet
  superconductor ${\mathrm{ute}}_{2}$},}\ }\href {\doibase
  10.1103/PhysRevB.100.220504} {\bibfield  {journal} {\bibinfo  {journal}
  {Phys. Rev. B}\ }\textbf {\bibinfo {volume} {100}},\ \bibinfo {pages}
  {220504} (\bibinfo {year} {2019})}\BibitemShut {NoStop}%
\bibitem [{\citenamefont {Jiao}\ \emph {et~al.}(2020)\citenamefont {Jiao},
  \citenamefont {Howard}, \citenamefont {Ran}, \citenamefont {Wang},
  \citenamefont {Rodriguez}, \citenamefont {Sigrist}, \citenamefont {Wang},
  \citenamefont {Butch},\ and\ \citenamefont {Madhavan}}]{Jiao2020}%
  \BibitemOpen
  \bibfield  {author} {\bibinfo {author} {\bibfnamefont {Lin}\ \bibnamefont
  {Jiao}}, \bibinfo {author} {\bibfnamefont {Sean}\ \bibnamefont {Howard}},
  \bibinfo {author} {\bibfnamefont {Sheng}\ \bibnamefont {Ran}}, \bibinfo
  {author} {\bibfnamefont {Zhenyu}\ \bibnamefont {Wang}}, \bibinfo {author}
  {\bibfnamefont {Jorge~Olivares}\ \bibnamefont {Rodriguez}}, \bibinfo {author}
  {\bibfnamefont {Manfred}\ \bibnamefont {Sigrist}}, \bibinfo {author}
  {\bibfnamefont {Ziqiang}\ \bibnamefont {Wang}}, \bibinfo {author}
  {\bibfnamefont {Nicholas~P.}\ \bibnamefont {Butch}}, \ and\ \bibinfo {author}
  {\bibfnamefont {Vidya}\ \bibnamefont {Madhavan}},\ }\bibfield  {title}
  {\enquote {\bibinfo {title} {Chiral superconductivity in heavy-fermion metal
  ute2},}\ }\href {\doibase 10.1038/s41586-020-2122-2} {\bibfield  {journal}
  {\bibinfo  {journal} {Nature}\ }\textbf {\bibinfo {volume} {579}},\ \bibinfo
  {pages} {523--527} (\bibinfo {year} {2020})}\BibitemShut {NoStop}%
\bibitem [{\citenamefont {Ishizuka}\ \emph {et~al.}(2019)\citenamefont
  {Ishizuka}, \citenamefont {Sumita}, \citenamefont {Daido},\ and\
  \citenamefont {Yanase}}]{PhysRevLett.123.217001}%
  \BibitemOpen
  \bibfield  {author} {\bibinfo {author} {\bibfnamefont {Jun}\ \bibnamefont
  {Ishizuka}}, \bibinfo {author} {\bibfnamefont {Shuntaro}\ \bibnamefont
  {Sumita}}, \bibinfo {author} {\bibfnamefont {Akito}\ \bibnamefont {Daido}}, \
  and\ \bibinfo {author} {\bibfnamefont {Youichi}\ \bibnamefont {Yanase}},\
  }\bibfield  {title} {\enquote {\bibinfo {title} {Insulator-metal transition
  and topological superconductivity in ${\mathrm{ute}}_{2}$ from a
  first-principles calculation},}\ }\href {\doibase
  10.1103/PhysRevLett.123.217001} {\bibfield  {journal} {\bibinfo  {journal}
  {Phys. Rev. Lett.}\ }\textbf {\bibinfo {volume} {123}},\ \bibinfo {pages}
  {217001} (\bibinfo {year} {2019})}\BibitemShut {NoStop}%
\bibitem [{\citenamefont {Braithwaite}\ \emph {et~al.}(2019)\citenamefont
  {Braithwaite}, \citenamefont {Vali{\v{s}}ka}, \citenamefont {Knebel},
  \citenamefont {Lapertot}, \citenamefont {Brison}, \citenamefont {Pourret},
  \citenamefont {Zhitomirsky}, \citenamefont {Flouquet}, \citenamefont
  {Honda},\ and\ \citenamefont {Aoki}}]{Braithwaite2019}%
  \BibitemOpen
  \bibfield  {author} {\bibinfo {author} {\bibfnamefont {D.}~\bibnamefont
  {Braithwaite}}, \bibinfo {author} {\bibfnamefont {M.}~\bibnamefont
  {Vali{\v{s}}ka}}, \bibinfo {author} {\bibfnamefont {G.}~\bibnamefont
  {Knebel}}, \bibinfo {author} {\bibfnamefont {G.}~\bibnamefont {Lapertot}},
  \bibinfo {author} {\bibfnamefont {J.-P.}\ \bibnamefont {Brison}}, \bibinfo
  {author} {\bibfnamefont {A.}~\bibnamefont {Pourret}}, \bibinfo {author}
  {\bibfnamefont {M.~E.}\ \bibnamefont {Zhitomirsky}}, \bibinfo {author}
  {\bibfnamefont {J.}~\bibnamefont {Flouquet}}, \bibinfo {author}
  {\bibfnamefont {F.}~\bibnamefont {Honda}}, \ and\ \bibinfo {author}
  {\bibfnamefont {D.}~\bibnamefont {Aoki}},\ }\bibfield  {title} {\enquote
  {\bibinfo {title} {Multiple superconducting phases in a nearly ferromagnetic
  system},}\ }\href {\doibase 10.1038/s42005-019-0248-z} {\bibfield  {journal}
  {\bibinfo  {journal} {Communications Physics}\ }\textbf {\bibinfo {volume}
  {2}},\ \bibinfo {pages} {147} (\bibinfo {year} {2019})}\BibitemShut {NoStop}%
\bibitem [{\citenamefont {Ran}\ \emph {et~al.}(2020)\citenamefont {Ran},
  \citenamefont {Kim}, \citenamefont {Liu}, \citenamefont {Saha}, \citenamefont
  {Hayes}, \citenamefont {Metz}, \citenamefont {Eo}, \citenamefont {Paglione},\
  and\ \citenamefont {Butch}}]{PhysRevB.101.140503}%
  \BibitemOpen
  \bibfield  {author} {\bibinfo {author} {\bibfnamefont {Sheng}\ \bibnamefont
  {Ran}}, \bibinfo {author} {\bibfnamefont {Hyunsoo}\ \bibnamefont {Kim}},
  \bibinfo {author} {\bibfnamefont {I-Lin}\ \bibnamefont {Liu}}, \bibinfo
  {author} {\bibfnamefont {Shanta~R.}\ \bibnamefont {Saha}}, \bibinfo {author}
  {\bibfnamefont {Ian}\ \bibnamefont {Hayes}}, \bibinfo {author} {\bibfnamefont
  {Tristin}\ \bibnamefont {Metz}}, \bibinfo {author} {\bibfnamefont {Yun~Suk}\
  \bibnamefont {Eo}}, \bibinfo {author} {\bibfnamefont {Johnpierre}\
  \bibnamefont {Paglione}}, \ and\ \bibinfo {author} {\bibfnamefont
  {Nicholas~P.}\ \bibnamefont {Butch}},\ }\bibfield  {title} {\enquote
  {\bibinfo {title} {Enhancement and reentrance of spin triplet
  superconductivity in ${\mathrm{ute}}_{2}$ under pressure},}\ }\href {\doibase
  10.1103/PhysRevB.101.140503} {\bibfield  {journal} {\bibinfo  {journal}
  {Phys. Rev. B}\ }\textbf {\bibinfo {volume} {101}},\ \bibinfo {pages}
  {140503} (\bibinfo {year} {2020})}\BibitemShut {NoStop}%
\bibitem [{\citenamefont {Bae}\ \emph {et~al.}(2019)\citenamefont {Bae},
  \citenamefont {Kim}, \citenamefont {Ran}, \citenamefont {Eo}, \citenamefont
  {Liu}, \citenamefont {Fuhrman}, \citenamefont {Paglione}, \citenamefont
  {Butch},\ and\ \citenamefont {Anlage}}]{bae2019anomalous}%
  \BibitemOpen
  \bibfield  {author} {\bibinfo {author} {\bibfnamefont {Seokjin}\ \bibnamefont
  {Bae}}, \bibinfo {author} {\bibfnamefont {Hyunsoo}\ \bibnamefont {Kim}},
  \bibinfo {author} {\bibfnamefont {Sheng}\ \bibnamefont {Ran}}, \bibinfo
  {author} {\bibfnamefont {Yun~Suk}\ \bibnamefont {Eo}}, \bibinfo {author}
  {\bibfnamefont {I-Lin}\ \bibnamefont {Liu}}, \bibinfo {author} {\bibfnamefont
  {Wesley}\ \bibnamefont {Fuhrman}}, \bibinfo {author} {\bibfnamefont
  {Johnpierre}\ \bibnamefont {Paglione}}, \bibinfo {author} {\bibfnamefont
  {Nicholas~P}\ \bibnamefont {Butch}}, \ and\ \bibinfo {author} {\bibfnamefont
  {Steven}\ \bibnamefont {Anlage}},\ }\href@noop {} {\enquote {\bibinfo {title}
  {Anomalous normal fluid response in a chiral superconductor},}\ } (\bibinfo
  {year} {2019}),\ \Eprint {http://arxiv.org/abs/1909.09032} {arXiv:1909.09032
  [cond-mat.supr-con]} \BibitemShut {NoStop}%
\bibitem [{\citenamefont {Hutanu}\ \emph {et~al.}(2019)\citenamefont {Hutanu},
  \citenamefont {Deng}, \citenamefont {Ran}, \citenamefont {Fuhrman},
  \citenamefont {Thoma},\ and\ \citenamefont {Butch}}]{hutanu2019crystal}%
  \BibitemOpen
  \bibfield  {author} {\bibinfo {author} {\bibfnamefont {V.}~\bibnamefont
  {Hutanu}}, \bibinfo {author} {\bibfnamefont {H.}~\bibnamefont {Deng}},
  \bibinfo {author} {\bibfnamefont {S.}~\bibnamefont {Ran}}, \bibinfo {author}
  {\bibfnamefont {W.~T.}\ \bibnamefont {Fuhrman}}, \bibinfo {author}
  {\bibfnamefont {H.}~\bibnamefont {Thoma}}, \ and\ \bibinfo {author}
  {\bibfnamefont {N.~P.}\ \bibnamefont {Butch}},\ }\href@noop {} {\enquote
  {\bibinfo {title} {Crystal structure of the unconventional spin-triplet
  superconductor ute2 at low temperature by single crystal neutron
  diffraction},}\ } (\bibinfo {year} {2019}),\ \Eprint
  {http://arxiv.org/abs/1905.04377} {arXiv:1905.04377 [cond-mat.supr-con]}
  \BibitemShut {NoStop}%
\bibitem [{\citenamefont {Sasaki}\ \emph {et~al.}(2003)\citenamefont {Sasaki},
  \citenamefont {Fukuda}, \citenamefont {Yoneyama},\ and\ \citenamefont
  {Kobayashi}}]{PhysRevB.67.144521}%
  \BibitemOpen
  \bibfield  {author} {\bibinfo {author} {\bibfnamefont {T.}~\bibnamefont
  {Sasaki}}, \bibinfo {author} {\bibfnamefont {T.}~\bibnamefont {Fukuda}},
  \bibinfo {author} {\bibfnamefont {N.}~\bibnamefont {Yoneyama}}, \ and\
  \bibinfo {author} {\bibfnamefont {N.}~\bibnamefont {Kobayashi}},\ }\bibfield
  {title} {\enquote {\bibinfo {title} {Shubnikov--de haas effect in the quantum
  vortex liquid state of the organic superconductor
  $\ensuremath{\kappa}\ensuremath{-}(\mathrm{BEDT}\ensuremath{-}\mathrm{TTF}{)}_{2}\mathrm{Cu}(\mathrm{NCS}{)}_{2}$},}\
  }\href {\doibase 10.1103/PhysRevB.67.144521} {\bibfield  {journal} {\bibinfo
  {journal} {Phys. Rev. B}\ }\textbf {\bibinfo {volume} {67}},\ \bibinfo
  {pages} {144521} (\bibinfo {year} {2003})}\BibitemShut {NoStop}%
\bibitem [{\citenamefont {Urano}\ \emph {et~al.}(2007)\citenamefont {Urano},
  \citenamefont {Tonishi}, \citenamefont {Inoue}, \citenamefont {Saito},
  \citenamefont {Fujiwara}, \citenamefont {Chiku}, \citenamefont {Oosawa},
  \citenamefont {Goto}, \citenamefont {Suzuki}, \citenamefont {Sasaki},
  \citenamefont {Kobayashi}, \citenamefont {Awaji},\ and\ \citenamefont
  {Watanabe}}]{PhysRevB.76.024505}%
  \BibitemOpen
  \bibfield  {author} {\bibinfo {author} {\bibfnamefont {M.}~\bibnamefont
  {Urano}}, \bibinfo {author} {\bibfnamefont {J.}~\bibnamefont {Tonishi}},
  \bibinfo {author} {\bibfnamefont {H.}~\bibnamefont {Inoue}}, \bibinfo
  {author} {\bibfnamefont {T.}~\bibnamefont {Saito}}, \bibinfo {author}
  {\bibfnamefont {T.}~\bibnamefont {Fujiwara}}, \bibinfo {author}
  {\bibfnamefont {H.}~\bibnamefont {Chiku}}, \bibinfo {author} {\bibfnamefont
  {A.}~\bibnamefont {Oosawa}}, \bibinfo {author} {\bibfnamefont
  {T.}~\bibnamefont {Goto}}, \bibinfo {author} {\bibfnamefont {T.}~\bibnamefont
  {Suzuki}}, \bibinfo {author} {\bibfnamefont {T.}~\bibnamefont {Sasaki}},
  \bibinfo {author} {\bibfnamefont {N.}~\bibnamefont {Kobayashi}}, \bibinfo
  {author} {\bibfnamefont {S.}~\bibnamefont {Awaji}}, \ and\ \bibinfo {author}
  {\bibfnamefont {K.}~\bibnamefont {Watanabe}},\ }\bibfield  {title} {\enquote
  {\bibinfo {title} {Nmr study of the vortex slush phase in organic
  superconductor
  $\ensuremath{\kappa}\text{\ensuremath{-}}(\mathrm{BEDT}\text{\ensuremath{-}}\mathrm{TTF}{)}_{2}\mathrm{Cu}(\mathrm{NCS}{)}_{2}$},}\
  }\href {\doibase 10.1103/PhysRevB.76.024505} {\bibfield  {journal} {\bibinfo
  {journal} {Phys. Rev. B}\ }\textbf {\bibinfo {volume} {76}},\ \bibinfo
  {pages} {024505} (\bibinfo {year} {2007})}\BibitemShut {NoStop}%
\bibitem [{\citenamefont {Sasaki}\ \emph {et~al.}(2002)\citenamefont {Sasaki},
  \citenamefont {Fukuda}, \citenamefont {Nishizaki}, \citenamefont {Fujita},
  \citenamefont {Yoneyama}, \citenamefont {Kobayashi},\ and\ \citenamefont
  {Biberacher}}]{PhysRevB.66.224513}%
  \BibitemOpen
  \bibfield  {author} {\bibinfo {author} {\bibfnamefont {T.}~\bibnamefont
  {Sasaki}}, \bibinfo {author} {\bibfnamefont {T.}~\bibnamefont {Fukuda}},
  \bibinfo {author} {\bibfnamefont {T.}~\bibnamefont {Nishizaki}}, \bibinfo
  {author} {\bibfnamefont {T.}~\bibnamefont {Fujita}}, \bibinfo {author}
  {\bibfnamefont {N.}~\bibnamefont {Yoneyama}}, \bibinfo {author}
  {\bibfnamefont {N.}~\bibnamefont {Kobayashi}}, \ and\ \bibinfo {author}
  {\bibfnamefont {W.}~\bibnamefont {Biberacher}},\ }\bibfield  {title}
  {\enquote {\bibinfo {title} {Low-temperature vortex liquid states induced by
  quantum fluctuations in the quasi-two-dimensional organic superconductor
  $\ensuremath{\kappa}\ensuremath{-}(\mathrm{BEDT}\ensuremath{-}\mathrm{TTF}{)}_{2}\mathrm{Cu}(\mathrm{NCS}{)}_{2}$},}\
  }\href {\doibase 10.1103/PhysRevB.66.224513} {\bibfield  {journal} {\bibinfo
  {journal} {Phys. Rev. B}\ }\textbf {\bibinfo {volume} {66}},\ \bibinfo
  {pages} {224513} (\bibinfo {year} {2002})}\BibitemShut {NoStop}%
\bibitem [{\citenamefont {Lortz}\ \emph {et~al.}(2007)\citenamefont {Lortz},
  \citenamefont {Wang}, \citenamefont {Demuer}, \citenamefont {B\"ottger},
  \citenamefont {Bergk}, \citenamefont {Zwicknagl}, \citenamefont {Nakazawa},\
  and\ \citenamefont {Wosnitza}}]{PhysRevLett.99.187002}%
  \BibitemOpen
  \bibfield  {author} {\bibinfo {author} {\bibfnamefont {R.}~\bibnamefont
  {Lortz}}, \bibinfo {author} {\bibfnamefont {Y.}~\bibnamefont {Wang}},
  \bibinfo {author} {\bibfnamefont {A.}~\bibnamefont {Demuer}}, \bibinfo
  {author} {\bibfnamefont {P.~H.~M.}\ \bibnamefont {B\"ottger}}, \bibinfo
  {author} {\bibfnamefont {B.}~\bibnamefont {Bergk}}, \bibinfo {author}
  {\bibfnamefont {G.}~\bibnamefont {Zwicknagl}}, \bibinfo {author}
  {\bibfnamefont {Y.}~\bibnamefont {Nakazawa}}, \ and\ \bibinfo {author}
  {\bibfnamefont {J.}~\bibnamefont {Wosnitza}},\ }\bibfield  {title} {\enquote
  {\bibinfo {title} {Calorimetric evidence for a
  fulde-ferrell-larkin-ovchinnikov superconducting state in the layered organic
  superconductor
  $\ensuremath{\kappa}\mathrm{\text{\ensuremath{-}}}(\mathrm{BEDT}\mathrm{\text{\ensuremath{-}}}\mathrm{TTF}{)}_{2}\mathrm{Cu}(\mathrm{NCS}{)}_{2}$},}\
  }\href {\doibase 10.1103/PhysRevLett.99.187002} {\bibfield  {journal}
  {\bibinfo  {journal} {Phys. Rev. Lett.}\ }\textbf {\bibinfo {volume} {99}},\
  \bibinfo {pages} {187002} (\bibinfo {year} {2007})}\BibitemShut {NoStop}%
\bibitem [{\citenamefont {Izawa}\ \emph {et~al.}(2001)\citenamefont {Izawa},
  \citenamefont {Yamaguchi}, \citenamefont {Sasaki},\ and\ \citenamefont
  {Matsuda}}]{PhysRevLett.88.027002}%
  \BibitemOpen
  \bibfield  {author} {\bibinfo {author} {\bibfnamefont {K.}~\bibnamefont
  {Izawa}}, \bibinfo {author} {\bibfnamefont {H.}~\bibnamefont {Yamaguchi}},
  \bibinfo {author} {\bibfnamefont {T.}~\bibnamefont {Sasaki}}, \ and\ \bibinfo
  {author} {\bibfnamefont {Yuji}\ \bibnamefont {Matsuda}},\ }\bibfield  {title}
  {\enquote {\bibinfo {title} {Superconducting gap structure of
  $\mathit{\ensuremath{\kappa}}\ensuremath{-}(\mathrm{BEDT}\ensuremath{-}\mathrm{TTF}{)}_{2}\mathrm{Cu}(\mathrm{NCS}{)}_{2}$
  probed by thermal conductivity tensor},}\ }\href {\doibase
  10.1103/PhysRevLett.88.027002} {\bibfield  {journal} {\bibinfo  {journal}
  {Phys. Rev. Lett.}\ }\textbf {\bibinfo {volume} {88}},\ \bibinfo {pages}
  {027002} (\bibinfo {year} {2001})}\BibitemShut {NoStop}%
\bibitem [{\citenamefont {Banerjee}\ \emph {et~al.}(2013)\citenamefont
  {Banerjee}, \citenamefont {Zhang},\ and\ \citenamefont
  {Randeria}}]{Banerjee2013}%
  \BibitemOpen
  \bibfield  {author} {\bibinfo {author} {\bibfnamefont {Sumilan}\ \bibnamefont
  {Banerjee}}, \bibinfo {author} {\bibfnamefont {Shizhong}\ \bibnamefont
  {Zhang}}, \ and\ \bibinfo {author} {\bibfnamefont {Mohit}\ \bibnamefont
  {Randeria}},\ }\bibfield  {title} {\enquote {\bibinfo {title} {Theory of
  quantum oscillations in the vortex-liquid state of high-tc
  superconductors},}\ }\href {\doibase 10.1038/ncomms2667} {\bibfield
  {journal} {\bibinfo  {journal} {Nature Communications}\ }\textbf {\bibinfo
  {volume} {4}},\ \bibinfo {pages} {1700} (\bibinfo {year} {2013})}\BibitemShut
  {NoStop}%
\bibitem [{\citenamefont {Song}\ and\ \citenamefont
  {Koshelev}(2017)}]{PhysRevB.95.174503}%
  \BibitemOpen
  \bibfield  {author} {\bibinfo {author} {\bibfnamefont {Kok~Wee}\ \bibnamefont
  {Song}}\ and\ \bibinfo {author} {\bibfnamefont {Alexei~E.}\ \bibnamefont
  {Koshelev}},\ }\bibfield  {title} {\enquote {\bibinfo {title} {Strong
  landau-quantization effects in high-magnetic-field superconductivity of a
  two-dimensional multiple-band metal near the lifshitz transition},}\ }\href
  {\doibase 10.1103/PhysRevB.95.174503} {\bibfield  {journal} {\bibinfo
  {journal} {Phys. Rev. B}\ }\textbf {\bibinfo {volume} {95}},\ \bibinfo
  {pages} {174503} (\bibinfo {year} {2017})}\BibitemShut {NoStop}%
\bibitem [{\citenamefont {Song}\ and\ \citenamefont
  {Koshelev}(2018)}]{PhysRevB.97.224520}%
  \BibitemOpen
  \bibfield  {author} {\bibinfo {author} {\bibfnamefont {Kok~Wee}\ \bibnamefont
  {Song}}\ and\ \bibinfo {author} {\bibfnamefont {Alexei~E.}\ \bibnamefont
  {Koshelev}},\ }\bibfield  {title} {\enquote {\bibinfo {title} {Interplay
  between orbital-quantization effects and the fulde-ferrell-larkin-ovchinnikov
  instability in multiple-band layered superconductors},}\ }\href {\doibase
  10.1103/PhysRevB.97.224520} {\bibfield  {journal} {\bibinfo  {journal} {Phys.
  Rev. B}\ }\textbf {\bibinfo {volume} {97}},\ \bibinfo {pages} {224520}
  (\bibinfo {year} {2018})}\BibitemShut {NoStop}%
\bibitem [{\citenamefont {Harper}(1955)}]{Harper_1955}%
  \BibitemOpen
  \bibfield  {author} {\bibinfo {author} {\bibfnamefont {P~G}\ \bibnamefont
  {Harper}},\ }\bibfield  {title} {\enquote {\bibinfo {title} {The general
  motion of conduction electrons in a uniform magnetic field, with application
  to the diamagnetism of metals},}\ }\href {\doibase
  10.1088/0370-1298/68/10/305} {\bibfield  {journal} {\bibinfo  {journal}
  {Proceedings of the Physical Society. Section A}\ }\textbf {\bibinfo {volume}
  {68}},\ \bibinfo {pages} {879--892} (\bibinfo {year} {1955})}\BibitemShut
  {NoStop}%
\bibitem [{\citenamefont {Koshino}\ \emph {et~al.}(2001)\citenamefont
  {Koshino}, \citenamefont {Aoki}, \citenamefont {Kuroki}, \citenamefont
  {Kagoshima},\ and\ \citenamefont {Osada}}]{PhysRevLett.86.1062}%
  \BibitemOpen
  \bibfield  {author} {\bibinfo {author} {\bibfnamefont {M.}~\bibnamefont
  {Koshino}}, \bibinfo {author} {\bibfnamefont {H.}~\bibnamefont {Aoki}},
  \bibinfo {author} {\bibfnamefont {K.}~\bibnamefont {Kuroki}}, \bibinfo
  {author} {\bibfnamefont {S.}~\bibnamefont {Kagoshima}}, \ and\ \bibinfo
  {author} {\bibfnamefont {T.}~\bibnamefont {Osada}},\ }\bibfield  {title}
  {\enquote {\bibinfo {title} {Hofstadter butterfly and integer quantum hall
  effect in three dimensions},}\ }\href {\doibase 10.1103/PhysRevLett.86.1062}
  {\bibfield  {journal} {\bibinfo  {journal} {Phys. Rev. Lett.}\ }\textbf
  {\bibinfo {volume} {86}},\ \bibinfo {pages} {1062--1065} (\bibinfo {year}
  {2001})}\BibitemShut {NoStop}%
\bibitem [{\citenamefont {Koshino}\ \emph {et~al.}(2002)\citenamefont
  {Koshino}, \citenamefont {Aoki}, \citenamefont {Osada}, \citenamefont
  {Kuroki},\ and\ \citenamefont {Kagoshima}}]{PhysRevB.65.045310}%
  \BibitemOpen
  \bibfield  {author} {\bibinfo {author} {\bibfnamefont {M.}~\bibnamefont
  {Koshino}}, \bibinfo {author} {\bibfnamefont {H.}~\bibnamefont {Aoki}},
  \bibinfo {author} {\bibfnamefont {T.}~\bibnamefont {Osada}}, \bibinfo
  {author} {\bibfnamefont {K.}~\bibnamefont {Kuroki}}, \ and\ \bibinfo {author}
  {\bibfnamefont {S.}~\bibnamefont {Kagoshima}},\ }\bibfield  {title} {\enquote
  {\bibinfo {title} {Phase diagram for the hofstadter butterfly and integer
  quantum hall effect in three dimensions},}\ }\href {\doibase
  10.1103/PhysRevB.65.045310} {\bibfield  {journal} {\bibinfo  {journal} {Phys.
  Rev. B}\ }\textbf {\bibinfo {volume} {65}},\ \bibinfo {pages} {045310}
  (\bibinfo {year} {2002})}\BibitemShut {NoStop}%
\bibitem [{\citenamefont {Rajagopal}\ and\ \citenamefont
  {Vasudevan}(1966)}]{RAJAGOPAL1966539}%
  \BibitemOpen
  \bibfield  {author} {\bibinfo {author} {\bibfnamefont {A.K.}\ \bibnamefont
  {Rajagopal}}\ and\ \bibinfo {author} {\bibfnamefont {R.}~\bibnamefont
  {Vasudevan}},\ }\bibfield  {title} {\enquote {\bibinfo {title} {De haas-van
  alphen oscillations in the critical temperature of type ii
  superconductors},}\ }\href {\doibase
  https://doi.org/10.1016/0031-9163(66)90396-9} {\bibfield  {journal} {\bibinfo
   {journal} {Physics Letters}\ }\textbf {\bibinfo {volume} {23}},\ \bibinfo
  {pages} {539 -- 540} (\bibinfo {year} {1966})}\BibitemShut {NoStop}%
\bibitem [{\citenamefont {Te\ifmmode \check{s}\else
  \v{s}\fi{}anovi\ifmmode~\acute{c}\else \'{c}\fi{}}\ \emph
  {et~al.}(1989)\citenamefont {Te\ifmmode \check{s}\else
  \v{s}\fi{}anovi\ifmmode~\acute{c}\else \'{c}\fi{}}, \citenamefont {Rasolt},\
  and\ \citenamefont {Xing}}]{PhysRevLett.63.2425}%
  \BibitemOpen
  \bibfield  {author} {\bibinfo {author} {\bibfnamefont {Zlatko}\ \bibnamefont
  {Te\ifmmode \check{s}\else \v{s}\fi{}anovi\ifmmode~\acute{c}\else
  \'{c}\fi{}}}, \bibinfo {author} {\bibfnamefont {Mark}\ \bibnamefont
  {Rasolt}}, \ and\ \bibinfo {author} {\bibfnamefont {Lei}\ \bibnamefont
  {Xing}},\ }\bibfield  {title} {\enquote {\bibinfo {title} {Quantum limit of a
  flux lattice: Superconductivity and magnetic field in a new relationship},}\
  }\href {\doibase 10.1103/PhysRevLett.63.2425} {\bibfield  {journal} {\bibinfo
   {journal} {Phys. Rev. Lett.}\ }\textbf {\bibinfo {volume} {63}},\ \bibinfo
  {pages} {2425--2428} (\bibinfo {year} {1989})}\BibitemShut {NoStop}%
\bibitem [{\citenamefont {Chandrasekhar}(1962)}]{doi:10.1063/1.1777362}%
  \BibitemOpen
  \bibfield  {author} {\bibinfo {author} {\bibfnamefont {B.~S.}\ \bibnamefont
  {Chandrasekhar}},\ }\bibfield  {title} {\enquote {\bibinfo {title} {A note on
  the maximum critical field of high‐field superconductors},}\ }\href
  {\doibase 10.1063/1.1777362} {\bibfield  {journal} {\bibinfo  {journal}
  {Applied Physics Letters}\ }\textbf {\bibinfo {volume} {1}},\ \bibinfo
  {pages} {7--8} (\bibinfo {year} {1962})},\ \Eprint
  {http://arxiv.org/abs/https://doi.org/10.1063/1.1777362}
  {https://doi.org/10.1063/1.1777362} \BibitemShut {NoStop}%
\bibitem [{\citenamefont {Clogston}(1962)}]{PhysRevLett.9.266}%
  \BibitemOpen
  \bibfield  {author} {\bibinfo {author} {\bibfnamefont {A.~M.}\ \bibnamefont
  {Clogston}},\ }\bibfield  {title} {\enquote {\bibinfo {title} {Upper limit
  for the critical field in hard superconductors},}\ }\href {\doibase
  10.1103/PhysRevLett.9.266} {\bibfield  {journal} {\bibinfo  {journal} {Phys.
  Rev. Lett.}\ }\textbf {\bibinfo {volume} {9}},\ \bibinfo {pages} {266--267}
  (\bibinfo {year} {1962})}\BibitemShut {NoStop}%
\bibitem [{\citenamefont {Peotta}\ and\ \citenamefont
  {T{\"o}rm{\"a}}(2015)}]{Peotta2015}%
  \BibitemOpen
  \bibfield  {author} {\bibinfo {author} {\bibfnamefont {Sebastiano}\
  \bibnamefont {Peotta}}\ and\ \bibinfo {author} {\bibfnamefont {P{\"a}ivi}\
  \bibnamefont {T{\"o}rm{\"a}}},\ }\bibfield  {title} {\enquote {\bibinfo
  {title} {Superfluidity in topologically nontrivial flat bands},}\ }\href
  {\doibase 10.1038/ncomms9944} {\bibfield  {journal} {\bibinfo  {journal}
  {Nature Communications}\ }\textbf {\bibinfo {volume} {6}},\ \bibinfo {pages}
  {8944} (\bibinfo {year} {2015})}\BibitemShut {NoStop}%
\bibitem [{\citenamefont {Xie}\ \emph {et~al.}(2020)\citenamefont {Xie},
  \citenamefont {Song}, \citenamefont {Lian},\ and\ \citenamefont
  {Bernevig}}]{PhysRevLett.124.167002}%
  \BibitemOpen
  \bibfield  {author} {\bibinfo {author} {\bibfnamefont {Fang}\ \bibnamefont
  {Xie}}, \bibinfo {author} {\bibfnamefont {Zhida}\ \bibnamefont {Song}},
  \bibinfo {author} {\bibfnamefont {Biao}\ \bibnamefont {Lian}}, \ and\
  \bibinfo {author} {\bibfnamefont {B.~Andrei}\ \bibnamefont {Bernevig}},\
  }\bibfield  {title} {\enquote {\bibinfo {title} {Topology-bounded superfluid
  weight in twisted bilayer graphene},}\ }\href {\doibase
  10.1103/PhysRevLett.124.167002} {\bibfield  {journal} {\bibinfo  {journal}
  {Phys. Rev. Lett.}\ }\textbf {\bibinfo {volume} {124}},\ \bibinfo {pages}
  {167002} (\bibinfo {year} {2020})}\BibitemShut {NoStop}%
\bibitem [{\citenamefont {Cheng}(2010)}]{cheng2010quantum}%
  \BibitemOpen
  \bibfield  {author} {\bibinfo {author} {\bibfnamefont {Ran}\ \bibnamefont
  {Cheng}},\ }\href@noop {} {\enquote {\bibinfo {title} {Quantum geometric
  tensor (fubini-study metric) in simple quantum system: A pedagogical
  introduction},}\ } (\bibinfo {year} {2010}),\ \Eprint
  {http://arxiv.org/abs/1012.1337} {arXiv:1012.1337 [quant-ph]} \BibitemShut
  {NoStop}%
\bibitem [{\citenamefont {Marzari}\ and\ \citenamefont
  {Vanderbilt}(1997)}]{PhysRevB.56.12847}%
  \BibitemOpen
  \bibfield  {author} {\bibinfo {author} {\bibfnamefont {Nicola}\ \bibnamefont
  {Marzari}}\ and\ \bibinfo {author} {\bibfnamefont {David}\ \bibnamefont
  {Vanderbilt}},\ }\bibfield  {title} {\enquote {\bibinfo {title} {Maximally
  localized generalized wannier functions for composite energy bands},}\ }\href
  {\doibase 10.1103/PhysRevB.56.12847} {\bibfield  {journal} {\bibinfo
  {journal} {Phys. Rev. B}\ }\textbf {\bibinfo {volume} {56}},\ \bibinfo
  {pages} {12847--12865} (\bibinfo {year} {1997})}\BibitemShut {NoStop}%
\bibitem [{\citenamefont {Marzari}\ \emph {et~al.}(2012)\citenamefont
  {Marzari}, \citenamefont {Mostofi}, \citenamefont {Yates}, \citenamefont
  {Souza},\ and\ \citenamefont {Vanderbilt}}]{RevModPhys.84.1419}%
  \BibitemOpen
  \bibfield  {author} {\bibinfo {author} {\bibfnamefont {Nicola}\ \bibnamefont
  {Marzari}}, \bibinfo {author} {\bibfnamefont {Arash~A.}\ \bibnamefont
  {Mostofi}}, \bibinfo {author} {\bibfnamefont {Jonathan~R.}\ \bibnamefont
  {Yates}}, \bibinfo {author} {\bibfnamefont {Ivo}\ \bibnamefont {Souza}}, \
  and\ \bibinfo {author} {\bibfnamefont {David}\ \bibnamefont {Vanderbilt}},\
  }\bibfield  {title} {\enquote {\bibinfo {title} {Maximally localized wannier
  functions: Theory and applications},}\ }\href {\doibase
  10.1103/RevModPhys.84.1419} {\bibfield  {journal} {\bibinfo  {journal} {Rev.
  Mod. Phys.}\ }\textbf {\bibinfo {volume} {84}},\ \bibinfo {pages}
  {1419--1475} (\bibinfo {year} {2012})}\BibitemShut {NoStop}%
\bibitem [{\citenamefont {Julku}\ \emph {et~al.}(2016)\citenamefont {Julku},
  \citenamefont {Peotta}, \citenamefont {Vanhala}, \citenamefont {Kim},\ and\
  \citenamefont {T\"orm\"a}}]{PhysRevLett.117.045303}%
  \BibitemOpen
  \bibfield  {author} {\bibinfo {author} {\bibfnamefont {Aleksi}\ \bibnamefont
  {Julku}}, \bibinfo {author} {\bibfnamefont {Sebastiano}\ \bibnamefont
  {Peotta}}, \bibinfo {author} {\bibfnamefont {Tuomas~I.}\ \bibnamefont
  {Vanhala}}, \bibinfo {author} {\bibfnamefont {Dong-Hee}\ \bibnamefont {Kim}},
  \ and\ \bibinfo {author} {\bibfnamefont {P\"aivi}\ \bibnamefont
  {T\"orm\"a}},\ }\bibfield  {title} {\enquote {\bibinfo {title} {Geometric
  origin of superfluidity in the lieb-lattice flat band},}\ }\href {\doibase
  10.1103/PhysRevLett.117.045303} {\bibfield  {journal} {\bibinfo  {journal}
  {Phys. Rev. Lett.}\ }\textbf {\bibinfo {volume} {117}},\ \bibinfo {pages}
  {045303} (\bibinfo {year} {2016})}\BibitemShut {NoStop}%
\bibitem [{\citenamefont {Liang}\ \emph {et~al.}(2017)\citenamefont {Liang},
  \citenamefont {Vanhala}, \citenamefont {Peotta}, \citenamefont {Siro},
  \citenamefont {Harju},\ and\ \citenamefont {T\"orm\"a}}]{PhysRevB.95.024515}%
  \BibitemOpen
  \bibfield  {author} {\bibinfo {author} {\bibfnamefont {Long}\ \bibnamefont
  {Liang}}, \bibinfo {author} {\bibfnamefont {Tuomas~I.}\ \bibnamefont
  {Vanhala}}, \bibinfo {author} {\bibfnamefont {Sebastiano}\ \bibnamefont
  {Peotta}}, \bibinfo {author} {\bibfnamefont {Topi}\ \bibnamefont {Siro}},
  \bibinfo {author} {\bibfnamefont {Ari}\ \bibnamefont {Harju}}, \ and\
  \bibinfo {author} {\bibfnamefont {P\"aivi}\ \bibnamefont {T\"orm\"a}},\
  }\bibfield  {title} {\enquote {\bibinfo {title} {Band geometry, berry
  curvature, and superfluid weight},}\ }\href {\doibase
  10.1103/PhysRevB.95.024515} {\bibfield  {journal} {\bibinfo  {journal} {Phys.
  Rev. B}\ }\textbf {\bibinfo {volume} {95}},\ \bibinfo {pages} {024515}
  (\bibinfo {year} {2017})}\BibitemShut {NoStop}%
\bibitem [{\citenamefont {Peltonen}\ \emph {et~al.}(2018)\citenamefont
  {Peltonen}, \citenamefont {Ojaj\"arvi},\ and\ \citenamefont
  {Heikkil\"a}}]{PhysRevB.98.220504}%
  \BibitemOpen
  \bibfield  {author} {\bibinfo {author} {\bibfnamefont {Teemu~J.}\
  \bibnamefont {Peltonen}}, \bibinfo {author} {\bibfnamefont {Risto}\
  \bibnamefont {Ojaj\"arvi}}, \ and\ \bibinfo {author} {\bibfnamefont
  {Tero~T.}\ \bibnamefont {Heikkil\"a}},\ }\bibfield  {title} {\enquote
  {\bibinfo {title} {Mean-field theory for superconductivity in twisted bilayer
  graphene},}\ }\href {\doibase 10.1103/PhysRevB.98.220504} {\bibfield
  {journal} {\bibinfo  {journal} {Phys. Rev. B}\ }\textbf {\bibinfo {volume}
  {98}},\ \bibinfo {pages} {220504} (\bibinfo {year} {2018})}\BibitemShut
  {NoStop}%
\bibitem [{\citenamefont {Tinkham}(2004)}]{tinkham2004introduction}%
  \BibitemOpen
  \bibfield  {author} {\bibinfo {author} {\bibfnamefont {Michael}\ \bibnamefont
  {Tinkham}},\ }\href@noop {} {\emph {\bibinfo {title} {Introduction to
  superconductivity}}}\ (\bibinfo  {publisher} {Courier Corporation},\ \bibinfo
  {year} {2004})\BibitemShut {NoStop}%
\bibitem [{\citenamefont {Murakami}\ and\ \citenamefont
  {Nagaosa}(2003)}]{PhysRevLett.90.057002}%
  \BibitemOpen
  \bibfield  {author} {\bibinfo {author} {\bibfnamefont {Shuichi}\ \bibnamefont
  {Murakami}}\ and\ \bibinfo {author} {\bibfnamefont {Naoto}\ \bibnamefont
  {Nagaosa}},\ }\bibfield  {title} {\enquote {\bibinfo {title} {Berry phase in
  magnetic superconductors},}\ }\href {\doibase 10.1103/PhysRevLett.90.057002}
  {\bibfield  {journal} {\bibinfo  {journal} {Phys. Rev. Lett.}\ }\textbf
  {\bibinfo {volume} {90}},\ \bibinfo {pages} {057002} (\bibinfo {year}
  {2003})}\BibitemShut {NoStop}%
\bibitem [{\citenamefont {Li}\ and\ \citenamefont
  {Haldane}(2018)}]{PhysRevLett.120.067003}%
  \BibitemOpen
  \bibfield  {author} {\bibinfo {author} {\bibfnamefont {Yi}~\bibnamefont
  {Li}}\ and\ \bibinfo {author} {\bibfnamefont {F.~D.~M.}\ \bibnamefont
  {Haldane}},\ }\bibfield  {title} {\enquote {\bibinfo {title} {Topological
  nodal cooper pairing in doped weyl metals},}\ }\href {\doibase
  10.1103/PhysRevLett.120.067003} {\bibfield  {journal} {\bibinfo  {journal}
  {Phys. Rev. Lett.}\ }\textbf {\bibinfo {volume} {120}},\ \bibinfo {pages}
  {067003} (\bibinfo {year} {2018})}\BibitemShut {NoStop}%
\bibitem [{\citenamefont {Young}\ and\ \citenamefont
  {Wieder}(2017)}]{PhysRevLett.118.186401}%
  \BibitemOpen
  \bibfield  {author} {\bibinfo {author} {\bibfnamefont {Steve~M.}\
  \bibnamefont {Young}}\ and\ \bibinfo {author} {\bibfnamefont {Benjamin~J.}\
  \bibnamefont {Wieder}},\ }\bibfield  {title} {\enquote {\bibinfo {title}
  {Filling-enforced magnetic dirac semimetals in two dimensions},}\ }\href
  {\doibase 10.1103/PhysRevLett.118.186401} {\bibfield  {journal} {\bibinfo
  {journal} {Phys. Rev. Lett.}\ }\textbf {\bibinfo {volume} {118}},\ \bibinfo
  {pages} {186401} (\bibinfo {year} {2017})}\BibitemShut {NoStop}%
\bibitem [{\citenamefont {Watanabe}\ \emph {et~al.}(2016)\citenamefont
  {Watanabe}, \citenamefont {Po}, \citenamefont {Zaletel},\ and\ \citenamefont
  {Vishwanath}}]{PhysRevLett.117.096404}%
  \BibitemOpen
  \bibfield  {author} {\bibinfo {author} {\bibfnamefont {Haruki}\ \bibnamefont
  {Watanabe}}, \bibinfo {author} {\bibfnamefont {Hoi~Chun}\ \bibnamefont {Po}},
  \bibinfo {author} {\bibfnamefont {Michael~P.}\ \bibnamefont {Zaletel}}, \
  and\ \bibinfo {author} {\bibfnamefont {Ashvin}\ \bibnamefont {Vishwanath}},\
  }\bibfield  {title} {\enquote {\bibinfo {title} {Filling-enforced gaplessness
  in band structures of the 230 space groups},}\ }\href {\doibase
  10.1103/PhysRevLett.117.096404} {\bibfield  {journal} {\bibinfo  {journal}
  {Phys. Rev. Lett.}\ }\textbf {\bibinfo {volume} {117}},\ \bibinfo {pages}
  {096404} (\bibinfo {year} {2016})}\BibitemShut {NoStop}%
\bibitem [{\citenamefont {Park}\ \emph {et~al.}(2018)\citenamefont {Park},
  \citenamefont {Raza}, \citenamefont {Gilbert},\ and\ \citenamefont
  {Teo}}]{PhysRevB.98.184514}%
  \BibitemOpen
  \bibfield  {author} {\bibinfo {author} {\bibfnamefont {Moon~Jip}\
  \bibnamefont {Park}}, \bibinfo {author} {\bibfnamefont {Syed}\ \bibnamefont
  {Raza}}, \bibinfo {author} {\bibfnamefont {Matthew~J.}\ \bibnamefont
  {Gilbert}}, \ and\ \bibinfo {author} {\bibfnamefont {Jeffrey C.~Y.}\
  \bibnamefont {Teo}},\ }\bibfield  {title} {\enquote {\bibinfo {title}
  {Coupled wire models of interacting dirac nodal superconductors},}\ }\href
  {\doibase 10.1103/PhysRevB.98.184514} {\bibfield  {journal} {\bibinfo
  {journal} {Phys. Rev. B}\ }\textbf {\bibinfo {volume} {98}},\ \bibinfo
  {pages} {184514} (\bibinfo {year} {2018})}\BibitemShut {NoStop}%
\bibitem [{\citenamefont {Ryu}\ \emph {et~al.}(2010)\citenamefont {Ryu},
  \citenamefont {Schnyder}, \citenamefont {Furusaki},\ and\ \citenamefont
  {Ludwig}}]{Ryu_2010}%
  \BibitemOpen
  \bibfield  {author} {\bibinfo {author} {\bibfnamefont {Shinsei}\ \bibnamefont
  {Ryu}}, \bibinfo {author} {\bibfnamefont {Andreas~P}\ \bibnamefont
  {Schnyder}}, \bibinfo {author} {\bibfnamefont {Akira}\ \bibnamefont
  {Furusaki}}, \ and\ \bibinfo {author} {\bibfnamefont {Andreas W~W}\
  \bibnamefont {Ludwig}},\ }\bibfield  {title} {\enquote {\bibinfo {title}
  {Topological insulators and superconductors: tenfold way and dimensional
  hierarchy},}\ }\href {\doibase 10.1088/1367-2630/12/6/065010} {\bibfield
  {journal} {\bibinfo  {journal} {New Journal of Physics}\ }\textbf {\bibinfo
  {volume} {12}},\ \bibinfo {pages} {065010} (\bibinfo {year}
  {2010})}\BibitemShut {NoStop}%
\bibitem [{\citenamefont {Teo}\ and\ \citenamefont
  {Kane}(2010)}]{PhysRevB.82.115120}%
  \BibitemOpen
  \bibfield  {author} {\bibinfo {author} {\bibfnamefont {Jeffrey C.~Y.}\
  \bibnamefont {Teo}}\ and\ \bibinfo {author} {\bibfnamefont {C.~L.}\
  \bibnamefont {Kane}},\ }\bibfield  {title} {\enquote {\bibinfo {title}
  {Topological defects and gapless modes in insulators and superconductors},}\
  }\href {\doibase 10.1103/PhysRevB.82.115120} {\bibfield  {journal} {\bibinfo
  {journal} {Phys. Rev. B}\ }\textbf {\bibinfo {volume} {82}},\ \bibinfo
  {pages} {115120} (\bibinfo {year} {2010})}\BibitemShut {NoStop}%
\bibitem [{\citenamefont {Kim}\ \emph {et~al.}(2016)\citenamefont {Kim},
  \citenamefont {Park},\ and\ \citenamefont {Gilbert}}]{PhysRevB.93.214511}%
  \BibitemOpen
  \bibfield  {author} {\bibinfo {author} {\bibfnamefont {Youngseok}\
  \bibnamefont {Kim}}, \bibinfo {author} {\bibfnamefont {Moon~Jip}\
  \bibnamefont {Park}}, \ and\ \bibinfo {author} {\bibfnamefont {Matthew~J.}\
  \bibnamefont {Gilbert}},\ }\bibfield  {title} {\enquote {\bibinfo {title}
  {Probing unconventional superconductivity in inversion-symmetric doped weyl
  semimetal},}\ }\href {\doibase 10.1103/PhysRevB.93.214511} {\bibfield
  {journal} {\bibinfo  {journal} {Phys. Rev. B}\ }\textbf {\bibinfo {volume}
  {93}},\ \bibinfo {pages} {214511} (\bibinfo {year} {2016})}\BibitemShut
  {NoStop}%
\bibitem [{\citenamefont {Pacholski}\ \emph {et~al.}(2018)\citenamefont
  {Pacholski}, \citenamefont {Beenakker},\ and\ \citenamefont
  {Adagideli}}]{PhysRevLett.121.037701}%
  \BibitemOpen
  \bibfield  {author} {\bibinfo {author} {\bibfnamefont {M.~J.}\ \bibnamefont
  {Pacholski}}, \bibinfo {author} {\bibfnamefont {C.~W.~J.}\ \bibnamefont
  {Beenakker}}, \ and\ \bibinfo {author} {\bibfnamefont {\ifmmode
  \dot{I}\else~\.{I}\fi{}.}\ \bibnamefont {Adagideli}},\ }\bibfield  {title}
  {\enquote {\bibinfo {title} {Topologically protected landau level in the
  vortex lattice of a weyl superconductor},}\ }\href {\doibase
  10.1103/PhysRevLett.121.037701} {\bibfield  {journal} {\bibinfo  {journal}
  {Phys. Rev. Lett.}\ }\textbf {\bibinfo {volume} {121}},\ \bibinfo {pages}
  {037701} (\bibinfo {year} {2018})}\BibitemShut {NoStop}%
\bibitem [{\citenamefont {Miao}\ \emph {et~al.}(2020)\citenamefont {Miao},
  \citenamefont {Liu}, \citenamefont {Xu}, \citenamefont {Kotta}, \citenamefont
  {Kang}, \citenamefont {Ran}, \citenamefont {Paglione}, \citenamefont
  {Kotliar}, \citenamefont {Butch}, \citenamefont {Denlinger},\ and\
  \citenamefont {Wray}}]{PhysRevLett.124.076401}%
  \BibitemOpen
  \bibfield  {author} {\bibinfo {author} {\bibfnamefont {Lin}\ \bibnamefont
  {Miao}}, \bibinfo {author} {\bibfnamefont {Shouzheng}\ \bibnamefont {Liu}},
  \bibinfo {author} {\bibfnamefont {Yishuai}\ \bibnamefont {Xu}}, \bibinfo
  {author} {\bibfnamefont {Erica~C.}\ \bibnamefont {Kotta}}, \bibinfo {author}
  {\bibfnamefont {Chang-Jong}\ \bibnamefont {Kang}}, \bibinfo {author}
  {\bibfnamefont {Sheng}\ \bibnamefont {Ran}}, \bibinfo {author} {\bibfnamefont
  {Johnpierre}\ \bibnamefont {Paglione}}, \bibinfo {author} {\bibfnamefont
  {Gabriel}\ \bibnamefont {Kotliar}}, \bibinfo {author} {\bibfnamefont
  {Nicholas~P.}\ \bibnamefont {Butch}}, \bibinfo {author} {\bibfnamefont
  {Jonathan~D.}\ \bibnamefont {Denlinger}}, \ and\ \bibinfo {author}
  {\bibfnamefont {L.~Andrew}\ \bibnamefont {Wray}},\ }\bibfield  {title}
  {\enquote {\bibinfo {title} {Low energy band structure and symmetries of
  ${\mathrm{ute}}_{2}$ from angle-resolved photoemission spectroscopy},}\
  }\href {\doibase 10.1103/PhysRevLett.124.076401} {\bibfield  {journal}
  {\bibinfo  {journal} {Phys. Rev. Lett.}\ }\textbf {\bibinfo {volume} {124}},\
  \bibinfo {pages} {076401} (\bibinfo {year} {2020})}\BibitemShut {NoStop}%
\bibitem [{\citenamefont {Xu}\ \emph {et~al.}(2019)\citenamefont {Xu},
  \citenamefont {Sheng},\ and\ \citenamefont {Yang}}]{PhysRevLett.123.217002}%
  \BibitemOpen
  \bibfield  {author} {\bibinfo {author} {\bibfnamefont {Yuanji}\ \bibnamefont
  {Xu}}, \bibinfo {author} {\bibfnamefont {Yutao}\ \bibnamefont {Sheng}}, \
  and\ \bibinfo {author} {\bibfnamefont {Yi-feng}\ \bibnamefont {Yang}},\
  }\bibfield  {title} {\enquote {\bibinfo {title} {Quasi-two-dimensional fermi
  surfaces and unitary spin-triplet pairing in the heavy fermion superconductor
  ${\mathrm{ute}}_{2}$},}\ }\href {\doibase 10.1103/PhysRevLett.123.217002}
  {\bibfield  {journal} {\bibinfo  {journal} {Phys. Rev. Lett.}\ }\textbf
  {\bibinfo {volume} {123}},\ \bibinfo {pages} {217002} (\bibinfo {year}
  {2019})}\BibitemShut {NoStop}%
\bibitem [{\citenamefont {Knafo}\ \emph {et~al.}(2020)\citenamefont {Knafo},
  \citenamefont {Nardone}, \citenamefont {Valiska}, \citenamefont {Zitouni},
  \citenamefont {Lapertot}, \citenamefont {Aoki}, \citenamefont {Knebel},\ and\
  \citenamefont {Braithwaite}}]{knafo2020comparison}%
  \BibitemOpen
  \bibfield  {author} {\bibinfo {author} {\bibfnamefont {W.}~\bibnamefont
  {Knafo}}, \bibinfo {author} {\bibfnamefont {M.}~\bibnamefont {Nardone}},
  \bibinfo {author} {\bibfnamefont {M.}~\bibnamefont {Valiska}}, \bibinfo
  {author} {\bibfnamefont {A.}~\bibnamefont {Zitouni}}, \bibinfo {author}
  {\bibfnamefont {G.}~\bibnamefont {Lapertot}}, \bibinfo {author}
  {\bibfnamefont {D.}~\bibnamefont {Aoki}}, \bibinfo {author} {\bibfnamefont
  {G.}~\bibnamefont {Knebel}}, \ and\ \bibinfo {author} {\bibfnamefont
  {D.}~\bibnamefont {Braithwaite}},\ }\href@noop {} {\enquote {\bibinfo {title}
  {Comparison of two superconducting phases induced by a magnetic field in
  ute2},}\ } (\bibinfo {year} {2020}),\ \Eprint
  {http://arxiv.org/abs/2007.06009} {arXiv:2007.06009 [cond-mat.supr-con]}
  \BibitemShut {NoStop}%
\bibitem [{\citenamefont {Aoki}\ \emph {et~al.}(2001)\citenamefont {Aoki},
  \citenamefont {Huxley}, \citenamefont {Ressouche}, \citenamefont
  {Braithwaite}, \citenamefont {Flouquet}, \citenamefont {Brison},
  \citenamefont {Lhotel},\ and\ \citenamefont {Paulsen}}]{Aoki2001}%
  \BibitemOpen
  \bibfield  {author} {\bibinfo {author} {\bibfnamefont {Dai}\ \bibnamefont
  {Aoki}}, \bibinfo {author} {\bibfnamefont {Andrew}\ \bibnamefont {Huxley}},
  \bibinfo {author} {\bibfnamefont {Eric}\ \bibnamefont {Ressouche}}, \bibinfo
  {author} {\bibfnamefont {Daniel}\ \bibnamefont {Braithwaite}}, \bibinfo
  {author} {\bibfnamefont {Jacques}\ \bibnamefont {Flouquet}}, \bibinfo
  {author} {\bibfnamefont {Jean-Pascal}\ \bibnamefont {Brison}}, \bibinfo
  {author} {\bibfnamefont {Elsa}\ \bibnamefont {Lhotel}}, \ and\ \bibinfo
  {author} {\bibfnamefont {Carley}\ \bibnamefont {Paulsen}},\ }\bibfield
  {title} {\enquote {\bibinfo {title} {Coexistence of superconductivity and
  ferromagnetism in urhge},}\ }\href {\doibase 10.1038/35098048} {\bibfield
  {journal} {\bibinfo  {journal} {Nature}\ }\textbf {\bibinfo {volume} {413}},\
  \bibinfo {pages} {613--616} (\bibinfo {year} {2001})}\BibitemShut {NoStop}%
\bibitem [{\citenamefont {Huy}\ \emph {et~al.}(2007)\citenamefont {Huy},
  \citenamefont {Gasparini}, \citenamefont {de~Nijs}, \citenamefont {Huang},
  \citenamefont {Klaasse}, \citenamefont {Gortenmulder}, \citenamefont
  {de~Visser}, \citenamefont {Hamann}, \citenamefont {G\"orlach},\ and\
  \citenamefont {L\"ohneysen}}]{PhysRevLett.99.067006}%
  \BibitemOpen
  \bibfield  {author} {\bibinfo {author} {\bibfnamefont {N.~T.}\ \bibnamefont
  {Huy}}, \bibinfo {author} {\bibfnamefont {A.}~\bibnamefont {Gasparini}},
  \bibinfo {author} {\bibfnamefont {D.~E.}\ \bibnamefont {de~Nijs}}, \bibinfo
  {author} {\bibfnamefont {Y.}~\bibnamefont {Huang}}, \bibinfo {author}
  {\bibfnamefont {J.~C.~P.}\ \bibnamefont {Klaasse}}, \bibinfo {author}
  {\bibfnamefont {T.}~\bibnamefont {Gortenmulder}}, \bibinfo {author}
  {\bibfnamefont {A.}~\bibnamefont {de~Visser}}, \bibinfo {author}
  {\bibfnamefont {A.}~\bibnamefont {Hamann}}, \bibinfo {author} {\bibfnamefont
  {T.}~\bibnamefont {G\"orlach}}, \ and\ \bibinfo {author} {\bibfnamefont
  {H.~v.}\ \bibnamefont {L\"ohneysen}},\ }\bibfield  {title} {\enquote
  {\bibinfo {title} {Superconductivity on the border of weak itinerant
  ferromagnetism in ucoge},}\ }\href {\doibase 10.1103/PhysRevLett.99.067006}
  {\bibfield  {journal} {\bibinfo  {journal} {Phys. Rev. Lett.}\ }\textbf
  {\bibinfo {volume} {99}},\ \bibinfo {pages} {067006} (\bibinfo {year}
  {2007})}\BibitemShut {NoStop}%
\bibitem [{\citenamefont {Saxena}\ \emph {et~al.}(2000)\citenamefont {Saxena},
  \citenamefont {Agarwal}, \citenamefont {Ahilan}, \citenamefont {Grosche},
  \citenamefont {Haselwimmer}, \citenamefont {Steiner}, \citenamefont {Pugh},
  \citenamefont {Walker}, \citenamefont {Julian}, \citenamefont {Monthoux},
  \citenamefont {Lonzarich}, \citenamefont {Huxley}, \citenamefont {Sheikin},
  \citenamefont {Braithwaite},\ and\ \citenamefont {Flouquet}}]{Saxena2000}%
  \BibitemOpen
  \bibfield  {author} {\bibinfo {author} {\bibfnamefont {S.~S.}\ \bibnamefont
  {Saxena}}, \bibinfo {author} {\bibfnamefont {P.}~\bibnamefont {Agarwal}},
  \bibinfo {author} {\bibfnamefont {K.}~\bibnamefont {Ahilan}}, \bibinfo
  {author} {\bibfnamefont {F.~M.}\ \bibnamefont {Grosche}}, \bibinfo {author}
  {\bibfnamefont {R.~K.~W.}\ \bibnamefont {Haselwimmer}}, \bibinfo {author}
  {\bibfnamefont {M.~J.}\ \bibnamefont {Steiner}}, \bibinfo {author}
  {\bibfnamefont {E.}~\bibnamefont {Pugh}}, \bibinfo {author} {\bibfnamefont
  {I.~R.}\ \bibnamefont {Walker}}, \bibinfo {author} {\bibfnamefont {S.~R.}\
  \bibnamefont {Julian}}, \bibinfo {author} {\bibfnamefont {P.}~\bibnamefont
  {Monthoux}}, \bibinfo {author} {\bibfnamefont {G.~G.}\ \bibnamefont
  {Lonzarich}}, \bibinfo {author} {\bibfnamefont {A.}~\bibnamefont {Huxley}},
  \bibinfo {author} {\bibfnamefont {I.}~\bibnamefont {Sheikin}}, \bibinfo
  {author} {\bibfnamefont {D.}~\bibnamefont {Braithwaite}}, \ and\ \bibinfo
  {author} {\bibfnamefont {J.}~\bibnamefont {Flouquet}},\ }\bibfield  {title}
  {\enquote {\bibinfo {title} {Superconductivity on the border of
  itinerant-electron ferromagnetism in uge2},}\ }\href {\doibase
  10.1038/35020500} {\bibfield  {journal} {\bibinfo  {journal} {Nature}\
  }\textbf {\bibinfo {volume} {406}},\ \bibinfo {pages} {587--592} (\bibinfo
  {year} {2000})}\BibitemShut {NoStop}%
\bibitem [{\citenamefont {Mineev}(2017)}]{Mineev_2017}%
  \BibitemOpen
  \bibfield  {author} {\bibinfo {author} {\bibfnamefont {V~P}\ \bibnamefont
  {Mineev}},\ }\bibfield  {title} {\enquote {\bibinfo {title}
  {Superconductivity in uranium ferromagnets},}\ }\href {\doibase
  10.3367/ufne.2016.04.037771} {\bibfield  {journal} {\bibinfo  {journal}
  {Physics-Uspekhi}\ }\textbf {\bibinfo {volume} {60}},\ \bibinfo {pages}
  {121--148} (\bibinfo {year} {2017})}\BibitemShut {NoStop}%
\bibitem [{\citenamefont {Meng}\ and\ \citenamefont
  {Balents}(2012)}]{PhysRevB.86.054504}%
  \BibitemOpen
  \bibfield  {author} {\bibinfo {author} {\bibfnamefont {Tobias}\ \bibnamefont
  {Meng}}\ and\ \bibinfo {author} {\bibfnamefont {Leon}\ \bibnamefont
  {Balents}},\ }\bibfield  {title} {\enquote {\bibinfo {title} {Weyl
  superconductors},}\ }\href {\doibase 10.1103/PhysRevB.86.054504} {\bibfield
  {journal} {\bibinfo  {journal} {Phys. Rev. B}\ }\textbf {\bibinfo {volume}
  {86}},\ \bibinfo {pages} {054504} (\bibinfo {year} {2012})}\BibitemShut
  {NoStop}%
\bibitem [{\citenamefont {Fukui}\ \emph {et~al.}(2005)\citenamefont {Fukui},
  \citenamefont {Hatsugai},\ and\ \citenamefont
  {Suzuki}}]{doi:10.1143/JPSJ.74.1674}%
  \BibitemOpen
  \bibfield  {author} {\bibinfo {author} {\bibfnamefont {Takahiro}\
  \bibnamefont {Fukui}}, \bibinfo {author} {\bibfnamefont {Yasuhiro}\
  \bibnamefont {Hatsugai}}, \ and\ \bibinfo {author} {\bibfnamefont {Hiroshi}\
  \bibnamefont {Suzuki}},\ }\bibfield  {title} {\enquote {\bibinfo {title}
  {Chern numbers in discretized brillouin zone: Efficient method of computing
  (spin) hall conductances},}\ }\href {\doibase 10.1143/JPSJ.74.1674}
  {\bibfield  {journal} {\bibinfo  {journal} {Journal of the Physical Society
  of Japan}\ }\textbf {\bibinfo {volume} {74}},\ \bibinfo {pages} {1674--1677}
  (\bibinfo {year} {2005})},\ \Eprint
  {http://arxiv.org/abs/https://doi.org/10.1143/JPSJ.74.1674}
  {https://doi.org/10.1143/JPSJ.74.1674} \BibitemShut {NoStop}%
\end{thebibliography}%

\clearpage
\newpage

\begin{widetext}
	
	\beginsupplement
	
\section*{Supplementary Material for ``3D Hofstadter Butterfly Superconductor''}
	
	\tableofcontents

\section{spectrum of 3D Hofstadter Model}\label{sec:Spectrum}

In this section, we study the energy spectrum of the three-dimensional Hofstadter model in detail. For the clarity, we start our discussion by writing Eq. \eqref{eq:h0} again here. 
\begin{eqnarray}
H(k_y,k_z)_{3D} &=& \sum_{\langle x,x' \rangle}\textrm{T}_x c^\dagger_{x'} c_x+V_{3D}(x,k_y,k_z)c^\dagger_{x} c_x,
\label{Eq:sh3d}
\\
\nonumber
V_{3D}(x,k_y,k_z) &=& 2\textrm{T}_y \cos(2\pi \Phi_z x+k_y)+2\textrm{T}_z \cos(2\pi \Phi_y x+k_z ).
\end{eqnarray}
where $\langle,\rangle$ represents the nearest neighbor hopping and $\Phi_{y(z)}$ is the flux penetrating the $xz \,(xy)$ planes of the unit cell respectively. Ignoring $\textrm{T}_z$ first, the application of $\Phi_{z}$ generates the two-dimensional Landau bands. The low Landau bands form the well-localized eigenstates at the minima of the potential $2\textrm{T}_y \cos(2\pi \Phi_z x+k_y)$. Fig. \ref{Fig:Supwave} (a) shows the wave function of the lowest Landau levels localized at the minima of the potential(red solid line) . The finite tunneling between the localized envelopes of the Landau level gives rise to the effective band width of the mini-Landau bands, $\textrm{T}_\textrm{eff}$. Now, if we turn on $\textrm{T}_z$, the Landau bands feel the effective potential $2T_z \cos(2\pi\Phi_\textrm{eff}n_x-\Phi_\textrm{eff}k_y+k_z)$ (red dashed line in Fig. \ref{Fig:Supwave}). As a result, we can write down the effective Hamiltonian of the Landau bands as, 
\begin{eqnarray}
H(k_y,k_z)_{\textrm{eff}} &=& \sum_{\langle x,x' \rangle}\textrm{T}_\textrm{eff} J^\dagger_{x'} J_x+V_{\textrm{eff}}(x,k_y,k_z) J^\dagger_{x} J_x,
\label{Eq:sh2d}
\\
\nonumber
V_{\textrm{eff}}(x,k_y,k_z) &=& 2T_z \cos(2\pi\Phi_\textrm{eff}n_x-\Phi_\textrm{eff}k_y+k_z).
\end{eqnarray}
where $\Phi_\textrm{eff}=\frac{\Phi_y}{\Phi_z}$ is the enhanced flux due to the formation of the superlattice and $J_x$ is the annihilation operator of the Landau level. Since $\Phi_\textrm{eff}$ is the function of the tilt angle, the Hofstadter butterfly spectrum can be realized even if the magnetic length scale does not reach the primitive lattice lengths. The correspondence between Eq. \eqref{Eq:sh3d} and \eqref{Eq:sh2d} is explicitly confirmed in the full tight-binding model calculation in Fig. \ref{Fig:Supwave} (b). 

\begin{figure}
	\includegraphics[scale=0.7]{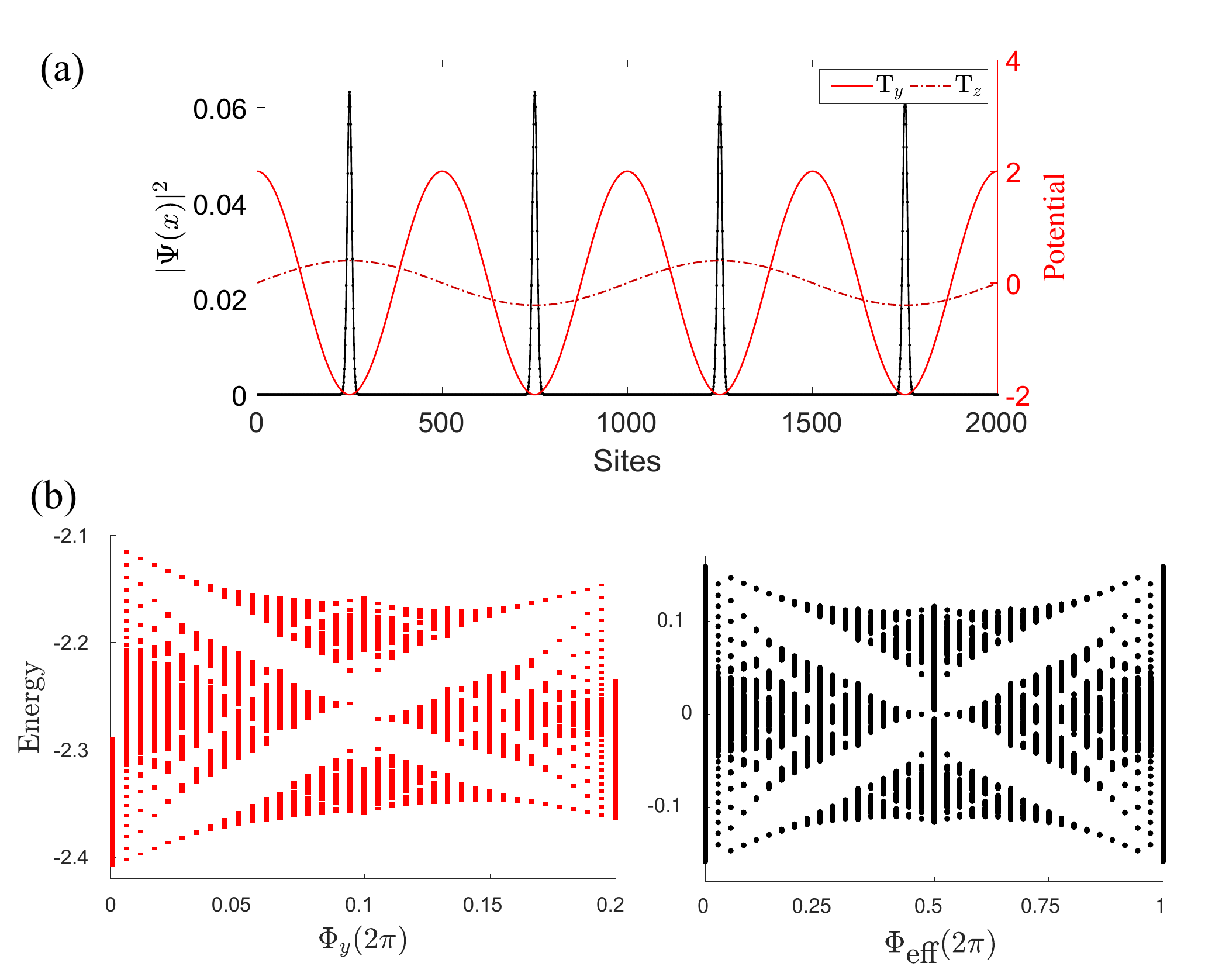}
	\caption{ (a) Illustration of the effective superlattice structure. The black line shows the wave functions of the four lowest Landau levels of Eq. \eqref{Eq:sh3d} with $\Phi_{z}=4/2000$, $\Phi_{y}=2/2000$. The wave functions are localized at the minima of the potential $\textrm{T}_y$(red solid line). The localized states are separated each other by $500$ lattice sites, which form the superlattice structure. the potential $T_z$(red dashed line) acting on the Landau level now effectively have the flux $\Phi_{\textrm{eff}}=\frac{1}{2}$. (b)-(c) Comparison of the energy spectrum of Eq. \eqref{Eq:sh3d} of the lowest Landau bands and Eq. \eqref{Eq:sh2d}. The application of the small tilted flux $\Phi_{y}$(left) shows the full Butterfly spectrum which has correspondence with the magnified effective flux $\Phi_{z}$(right). We set $\textrm{T}_x=1$,$\textrm{T}_y=0.5$,$\textrm{T}_z=0.005$. }
	\label{Fig:Supwave}
\end{figure}
\section{Superconducting Kernel Matrix}\label{sec:Kernel}

In this section, we describe the detailed numerical implementation of the superconducting Kernel matrix. To start our discussion, we would like to define the superconducting kernel matrix in a general setting. We define the Matusbara Green function of the full BdG Hamiltonian for imaginary time, $\tau$, as,
\begin{eqnarray}
G_{ij}(\tau)\equiv- \langle T_\tau c_i (\tau) c^\dagger_j (0) \rangle=\frac{1}{\beta}\sum_{\omega_n\in \mathbb{Z}}e^{-i\omega_n\tau}G_{ij}(i\omega_n),
\end{eqnarray}
where $\beta=1/ T$ and $i,j$ represents the orbital index. $\omega_n=(2n+1)\pi/\beta$ is the Matsubara frequency($n\in \mathbb{Z}$). We note that the anomalous Green function is not separately defined. $T_\tau$ is the imaginary time-ordering such that,
\begin{eqnarray}
T_\tau c_i (\tau) c^\dagger_j (0) \equiv 
\begin{cases}
	c_i (\tau) c^\dagger_j (0) \quad \textrm{if} \quad \tau >0 \\
	-c^\dagger_j (0) c_i (\tau)   \quad \textrm{if} \quad \tau <0.
\end{cases}
\end{eqnarray}

The equal-time correlation function can be written in terms of the Green function in the frequency space as,

\begin{eqnarray}
\langle  c^\dagger_j (0) c_i(0) \rangle= G_{ij}(\tau\rightarrow 0^-)=T\sum_{\omega_n\in \mathbb{Z}}G_{ij}(i\omega_n).
\label{eq:SOP}
\end{eqnarray}
We now aim to evaluate the Green function in the frequency space. In the Nambu basis, we can decompose the BdG Hamiltonian into the normal and the pairing parts respectively.
\begin{eqnarray}
H_{\textrm{BdG}}\equiv H_0(\mathbf{k})+H_{\textrm{pairing}}(\mathbf{k})=
\begin{pmatrix}
	h_0 (\mathbf{k})&0\\
	0&-h_0^T(-\mathbf{k})\\
\end{pmatrix}
+
\begin{pmatrix}
	0 & \Delta(\mathbf{k})\\
	\Delta^\dagger(\mathbf{k})&0\\
\end{pmatrix},
\end{eqnarray}
where $h_0(\mathbf{k})$ and $\Delta(\mathbf{k})$ are the normal state Hamiltonian and the pairing potential with the momentum $\mathbf{k}$ respectively. To further elaborate the above expansion, we conceive that $G_0(i\omega_n)$ has the following matrix structure,
\begin{eqnarray}
& G_0(\mathbf{k},\mathbf{k'},i\omega_n) =\delta_{\mathbf{k},\mathbf{k}'}G_0(\mathbf{k},i\omega_n)
\nonumber
\\
&=
\begin{pmatrix}
	\frac{1}{i\omega_n -h_0(\mathbf{k})}&0\\
	0&\frac{1}{i\omega_n +h_0^T(-\mathbf{k})}\\
\end{pmatrix}
\equiv
\begin{pmatrix}
	g_0(\mathbf{k},i\omega_n)&0\\
	0&-g_0^T(-\mathbf{k},-i\omega_n)\\
\end{pmatrix}.
\label{eq:Sg0}
\end{eqnarray}

The Green function of the BdG Hamiltonian can be decomposed into the normal part and the pairing part as well,
\begin{eqnarray}
\nonumber
G(i\omega_n)
&=&\frac{1}{i\omega_n-(H_0+H_{\textrm{pairing}}) }
\\
\nonumber
&=&\frac{1}{G_0^{-1}(i\omega_n)(I-G_0(i\omega_n)H_{\textrm{pairing}}) }
\\&=&\frac{1}{I-G_0(i\omega_n)H_{\textrm{pairing}}}G_0(i\omega_n),
\end{eqnarray}
where $G_0(i\omega_n)\equiv(i\omega_n-H_0)^{-1}$ is the Matsubara Green function of $H_0$. In the weak pairing limit, we can expand the above expression, using the relation, $(\frac{1}{1-M}\approx 1+M)$, as,
\begin{eqnarray}
\nonumber
G(i\omega_n)&\approx &( I+G_0(i\omega_n)H_{\textrm{pairing}})G_0(i\omega_n)
\\
&=& G_0(i\omega_n)+G_0(i\omega_n)H_{pairing}G_0(i\omega_n).
\label{eq:SG01}
\end{eqnarray}

The second term in Eq. \eqref{eq:SG01} is the first-order correction from the $H_{\textrm{pairing}}$ to the Green function.
By plugging Eq. \eqref{eq:SG01} to Eq. \eqref{eq:Sg0}, the first-order correction is given as, 
\begin{eqnarray}
\nonumber
G_0(i\omega_n)H_{pair}G_0(i\omega_n)
&=&
\begin{pmatrix}
	g_0(\mathbf{k},i\omega_n)&0\\
	0&-g_0^T(-\mathbf{k},-i\omega_n)\\
\end{pmatrix}
\begin{pmatrix}
	0&\Delta(\mathbf{k})\\
	\Delta(\mathbf{k})^\dagger&0\\
\end{pmatrix}
\begin{pmatrix}
	g_0(\mathbf{k},i\omega_n)&0\\
	0&-g_0^T(-\mathbf{k},-i\omega_n)\\
\end{pmatrix}
\\
&=&\begin{pmatrix}
	0&-g_0(\mathbf{k},i\omega_n)\Delta(\mathbf{k})g_0^T(-\mathbf{k},-i\omega_n)\\
	-g_0^T(-\mathbf{k},-i\omega_n)\Delta^\dagger(\mathbf{k})g_0(\mathbf{k},i\omega_n) &0\\
\end{pmatrix}.
\end{eqnarray}

From the above result, we can evaluate the equal-time correlation function in Eq. \eqref{eq:SOP} as, 
\begin{eqnarray}
\langle c_i^\dagger (\mathbf{k})c^\dagger_j (-\mathbf{k}) \rangle
\equiv
T\sum_{\omega_n}G_{i,(j+N_{\textrm{orb}})}(\mathbf{k},i\omega_n)=
-T\sum_{\omega_n}[g_0(\mathbf{k},i\omega_n)\Delta(\mathbf{k})g_0^T(-\mathbf{k},-i\omega_n)]_{ij}
\label{Eq:Stwopoint}
\end{eqnarray}
where $N_{\textrm{orb}}$ is the size of the matrix $h_0$. To derive the self-consistent gap equation, we also equate the left hand side of the equation to the superconducting order parameter. To do so, we consider the general form of the interaction as,
\begin{eqnarray}
H_{\textrm{int}}=\frac{1}{N}\sum_{\mathbf{k},\mathbf{k}',\mathbf{q}}V_{\alpha\beta\gamma\delta}(\mathbf{q})c^\dagger_\alpha(\mathbf{k+q}) c^\dagger_\beta(\mathbf{k'-q}) c_\gamma(\mathbf{k}') c_\delta(\mathbf{k}),
\label{Eq:SHI}
\end{eqnarray}
where $N=N_x N_y N_z$ is the number of the total three-dimensional lattice sites. Taking only the BCS channel where the center of the mass momentum is zero($\mathbf{k+k'}=0$), Eq. \eqref{Eq:SHI} reduces as,
\begin{eqnarray}
H_{\textrm{BCS}}=\frac{1}{N}\sum_{\mathbf{k},\mathbf{k}'} V_{\alpha\beta\gamma\delta}(\mathbf{k'-k})c^\dagger_\alpha(\mathbf{k'}) c^\dagger_\beta(\mathbf{-k'}) c_\gamma(-\mathbf{k}) c_\delta(\mathbf{k}).
\label{Eq:SHBCS}
\end{eqnarray} 

We perform the BCS mean-field decomposition of the above Hamiltonian as,
\begin{eqnarray}
H_{\textrm{pairing}}&=&
\frac{1}{N}\sum_{\mathbf{k},\mathbf{k}'}
V_{\alpha\beta\gamma\delta}(\mathbf{k'-k})c^\dagger_\alpha(\mathbf{k'}) c^\dagger_\beta(\mathbf{-k'}) c_\gamma(-\mathbf{k}) c_\delta(\mathbf{k})
\\
\nonumber
&=&
\frac{1}{N}\sum_{\mathbf{k},\mathbf{k}'}V_{\alpha\beta\gamma\delta}(\mathbf{k'-k})
[\langle c^\dagger_\alpha(\mathbf{k'}) c^\dagger_\beta(\mathbf{-k'}) \rangle c_\gamma(-\mathbf{k}) c_\delta(\mathbf{k})
+
c^\dagger_\alpha(\mathbf{k'}) c^\dagger_\beta(\mathbf{-k'}) \langle c_\gamma(-\mathbf{k}) c_\delta(\mathbf{k}) \rangle]
\\
\nonumber
&+&
\frac{1}{N}\sum_{\mathbf{k},\mathbf{k}'}V_{\alpha\beta\gamma\delta}(\mathbf{k'-k})
[\overline{c^\dagger_\alpha(\mathbf{k'})c^\dagger_\beta(\mathbf{-k'})}
\,
\overline{c_\gamma(\mathbf{k})c_\delta(\mathbf{-k})}- \langle c^\dagger_\alpha(\mathbf{k'})c^\dagger_\beta(\mathbf{-k'})\rangle \langle c_\gamma(\mathbf{k})c_\delta(\mathbf{-k})\rangle]
\\
&\approx&\sum_{\mathbf{k}}\Delta_{\gamma\delta}(\mathbf{k})
c_\gamma(-\mathbf{k}) c_\delta(\mathbf{k})
+
\Delta^*_{\alpha\beta}(\mathbf{k})
c^\dagger_\alpha(\mathbf{k}) c^\dagger_\beta(\mathbf{-k})+\textrm{constants}.
\nonumber
\end{eqnarray}
where $\overline{O}\equiv \hat{O}-\langle \hat{O} \rangle $ indicates the fluctuation. Using Eq. \eqref{Eq:Stwopoint}, we derive the generic linearized gap equation for the multi-band system as following,
\begin{eqnarray}
\nonumber
\Delta_{\gamma\delta}(\mathbf{k})=\sum_{\mathbf{k}',\alpha\beta}V_{\alpha\beta\gamma\delta}(\mathbf{k'-k}) \langle c^\dagger_\alpha(\mathbf{k}') c^\dagger_\beta(-\mathbf{k}') \rangle
\\
=-T\sum_{k',\alpha\beta,\omega_n}V_{\alpha\beta\gamma\delta}(\mathbf{k'-k})  [g_0(\mathbf{k}',i\omega_n)\Delta(\mathbf{k'})g_0^T(-\mathbf{k}',-i\omega_n)]_{\alpha,\beta}
\nonumber
\\
=\bigg[ -T\sum_{k',\alpha\beta,\omega_n}V_{\alpha\beta\gamma\delta}(\mathbf{k'-k})  [ g_{0}(\mathbf{k'},i\omega_n) ]_{\alpha \lambda}
[g_{0}(-\mathbf{k},-i\omega_n)]_{\beta \mu} \bigg]
\Delta_{\lambda \mu}(\mathbf{k'}).
\label{Eq:Sself}
\end{eqnarray}
We notice that the above equation is a form of tensorial multiplication, and it can be simplified by introducing the superconducting pairing kernel, which is given as,
\begin{eqnarray}
K_{\gamma\delta,\lambda\mu}(\mathbf{k},\mathbf{k}')=-T\sum_{k,\omega_n}V_{\alpha\beta\gamma\delta}(\mathbf{k'-k})  [ g_{0}(\mathbf{k'},i\omega_n) ]_{\alpha \lambda}
[g_{0}(-\mathbf{k},-i\omega_n)]_{\beta \mu}
\end{eqnarray}
The linearized self-consistent equation in Eq. \eqref{Eq:Sself} reduces to
\begin{eqnarray}
\Delta_{\gamma\delta}(\mathbf{k})=\sum_{\mathbf{k}',\lambda\mu}K_{\gamma\delta,\lambda\mu}(\mathbf{k},\mathbf{k}')\Delta_{\lambda \mu}(\mathbf{k'})
\end{eqnarray}
We numerically diagonalize the Kernel matrix and derive the eignevalues $\lambda_n$ as a function of the temperature. The onset of the superconducting instability is indicated by the condition, $\textrm{max}[\lambda_n](T)\ge 1$. The superconducting critical temperature is derived by the condition, $\textrm{max}[\lambda_n](T_c)= 1$.

\section{Quantum Geometric Tensor}\label{sec:QGT}

Before going to the details of the superfluid weight, we first introduce the quantum geometric tensor(QGT) and several generic properties. We mainly follow the discussion of Ref. \cite{cheng2010quantum}. For the generic Bloch Hamiltonian $H(\mathbf{k})$ with the momentum $\mathbf{k}$, we can diagonalize the Hamiltonian as,
\begin{eqnarray}
H(\mathbf{k})=U(\mathbf{k})D(\mathbf{k})U^\dagger((\mathbf{k}),
\label{eq:Sdiag}
\end{eqnarray}
where $D=\textrm{diag}(\epsilon_{1},\epsilon_{2},...\epsilon_{n})$ is the diagonal matrix containing energies. $U(\mathbf{k})=(u_{1}(\mathbf{k}),u_{2}(\mathbf{k}),...u_{n}(\mathbf{k}))$ is the unitary matrix, where the column vectors are the eigenstates. Then, we define the QGT of the $i$-th band as,

\begin{eqnarray}
\nonumber
g_{\mu\nu}(\mathbf{k})&\equiv& \partial_{\mu}u^\dagger_i (\mathbf{k})\partial_{\nu}u_i (\mathbf{k})
-[\partial_{\mu}u^\dagger_i (\mathbf{k})u_i(\mathbf{k})][
u^\dagger_i (\mathbf{k}) \partial_\nu u_i (\mathbf{k}) ]=g_{\nu\mu}^*
\\
&=& \sum_{j\neq i}\partial_{\mu} u_i^\dagger (\mathbf{k}) u_j (\mathbf{k}) u_j^\dagger(\mathbf{k}) \partial_{\nu}u_i (\mathbf{k}).
\label{eq:Sgdef}
\end{eqnarray}

Furthermore, the QGT can be decomposed into the real and imaginary parts as,
\begin{eqnarray}
\textrm{Re}g_{\mu\nu}(\mathbf{k})
&=&
\frac{1}{2}(g_{\mu\nu}+g_{\nu\mu})
\nonumber
\\
\label{eq:sReQ}
&=&
\frac{1}{2}
(\partial_{\mu}u^\dagger_i (\mathbf{k})\partial_{\nu}u_i (\mathbf{k})
+\partial_{\nu}u^\dagger_i (\mathbf{k})\partial_{\mu}u_i (\mathbf{k})
+2[u^\dagger_i (\mathbf{k})\partial_{\mu}u_i(\mathbf{k})][
u^\dagger_i (\mathbf{k}) \partial_\nu u_i (\mathbf{k}) ])
\\
\textrm{Im}g_{\mu\nu}(\mathbf{k})&=&
\frac{1}{2i}(g_{\mu\nu}-g_{\nu\mu})
\nonumber
\\
&=& -\frac{i}{2} (\partial_{\mu}u_i^\dagger (\mathbf{k})\partial_{\nu}u_i (\mathbf{k})-\partial_{\nu}u_i^\dagger (\mathbf{k})\partial_{\mu}u_i (\mathbf{k}))=-\frac{1}{2}F_{\mu\nu}
\end{eqnarray}
where $F_{\mu\nu}$ is the Berry curvature. An important property of the QGT is the positive semidefiniteness such that, for a complex vector $\mathbf{c}$, the following relations is satisfied,
\begin{eqnarray}
\sum_{\mu,\nu=x,y}c^\dagger_\mu g_{\mu\nu}(\mathbf{k}) c_\nu \ge 0.
\end{eqnarray}
If we choose $\mathbf{c}=(1,i)$, we derive the relation that
\begin{eqnarray}
g_{xx}(\mathbf{k})+g_{yy}(\mathbf{k})+i(g_{xy}(\mathbf{k})-g_{yx}(\mathbf{k}))\ge 0.
\end{eqnarray}
In the isotropic system($T_x=T_y$), we find that $g_{xx,yy}(\mathbf{k})\ge \textrm{Im} g_{xy}(\mathbf{k})$.

%

\section{Calculation of Superfluid Weights}\label{sec:Ds}
In this section, we now explain the relationship between the superfluid weigth and the QGT of the Hofstadter butterfly. We consider the following decomposition of the generic BdG Hamiltonian,
\begin{eqnarray}
\nonumber
H_{\textrm{BdG}}&=&\begin{pmatrix}
	h_0(\mk) & \Delta(\mk)\\
	\Delta^\dagger(\mk) & -h_0^T(-\mk)\\
\end{pmatrix}
=
\begin{pmatrix}
	U(\mk) D(\mk) U^\dagger(\mk) & \Delta(\mk)\\
	\Delta^\dagger(\mk) & -U^*({-\mk})D({-\mk})U^T({-\mk})\\
\end{pmatrix}
\\
&=&
\begin{pmatrix}
	U(\mk) & \\
	& U^*(-\mk)\\
\end{pmatrix}
\begin{pmatrix}
	D(\mk) & U^\dagger(\mk)\Delta(\mk)U^*({-\mk})\\
	U^T(-\mk)\Delta^\dagger(\mk)U(\mk) & -D({-\mk})\\
\end{pmatrix}
\begin{pmatrix}
	U^\dagger(\mk) & \\
	& U^T(-\mk)\\
\end{pmatrix}.
\end{eqnarray}
Here we have utilized the diagonalization in Eq. \eqref{eq:Sdiag}. If the Fermi level is near $n$-th band and the interband gap is sufficiently larger than the interaction strength, we can project the Hamiltonian to the $n$-th band. In such cases, we can generally write the form of the unitary matrix and the order parameter matrix as,
\begin{eqnarray}
[U^\dagger(\mk)\Delta(\mk)U^*({-\mk})]_{i,j}=\delta_{in}\delta_{jn}\Delta_{\textrm{proj}}(\mathbf{k}),
\quad\quad
[\Delta(\mk)]_{i,j}=[U(\mk)]_{i,n}[U^T(-\mk)]_{n,j} \Delta_{\textrm{proj}}(\mathbf{k})
\label{Eq:Sdecomp}
\end{eqnarray}
where $\Delta_{\textrm{proj}}(\mathbf{k})$ is a complex order parameter acting on the projected band. We now consider the case where the non-zero supercurrent flows with the superconducting order parameter gradient, $\Delta(\mathbf{r})=|\Delta|e^{2i \mathbf{q}\cdot \mathbf{r}}$. The corresponding BdG Hamiltonian couples the electron sector of $\mathbf{k+q}$ and the hole sector of $\mathbf{-k+q}$ like the case of the Fulde-Ferrell states. The Hamiltonian is now written as, 
\begin{eqnarray}
\nonumber
H_{\textrm{BdG}}(\mathbf{q})
\equiv
\begin{pmatrix}
	h_0(\mathbf{k+q}) & \Delta(\mk)\\
	\Delta^\dagger(\mk) & -h_0^T(\mathbf{-k+q})\\
\end{pmatrix}.
\end{eqnarray}
We now take the similar decomposition in Eq. \eqref{Eq:Sdecomp},

\begin{eqnarray}
H_{\textrm{BdG}}(\mathbf{q})=
\begin{pmatrix}
	U(\mathbf{k+q})D(\mathbf{k+q})U^\dagger(\mathbf{k-q}) & \Delta(\mathbf{k})\\
	\Delta^\dagger(\mathbf{k}) & -U^*(\mathbf{-k+q})D(\mathbf{-k+q})U^T(\mathbf{-k+q})\\
\end{pmatrix}
\end{eqnarray}
\begin{eqnarray}
=
\mathcal{U}(\mk,\mathbf{q})
\begin{pmatrix}
	D(\mathbf{k+q}) & U^\dagger(\mathbf{k+q})\Delta(\mk)U^*(\mathbf{-k+q})\\
	U^T(\mathbf{-k+q})\Delta^\dagger(\mk)U(\mathbf{k+q}) & -D(\mathbf{-k+q})\\
\end{pmatrix}
\mathcal{U}(\mk,\mathbf{q})^\dagger ,
\end{eqnarray}
where $\mathcal{U}(\mk,\mathbf{q})\equiv\textrm{diag}(U(\mathbf{k+q}),U^*(\mathbf{-k+q}))$. Now the projected Hamiltonian on the $n$-th band can be written as,
\begin{eqnarray}
H_{\textrm{proj},nn}(\mathbf{q})\approx \begin{pmatrix}
	\epsilon(\mathbf{k+q}) & \Delta_{\textrm{proj}}(\mk,\mathbf{q})\\
	\Delta_{\textrm{proj}}^\dagger(\mk,\mathbf{q}) & -\epsilon(\mathbf{-k+q})\\
\end{pmatrix}
\end{eqnarray}
where $\epsilon(\mk)$ is the normal state energy of the projected band. The expression of $\Delta_{\textrm{proj}}$ can be derived using Eq. \eqref{Eq:Sdecomp},
\begin{eqnarray}
\nonumber
\Delta_{\textrm{proj}}(\mk,\mathbf{q})
&=&
 U^\dagger(\mathbf{k+q})\Delta(\mk)U^*(\mathbf{-k+q})
\\
&=&
\sum_{N_1,N_2}[U^\dagger(\mathbf{k+q})]_{n,N_1}[U(\mk)]_{N_1,n}[U^T(\mathbf{-k})]_{n,N_2} [U^*(\mathbf{-k+q})]_{N_2,n}\Delta_{\textrm{proj}}(\mathbf{k}).
\end{eqnarray}
We have now derive the generic expression of the superconducting order parameter with the finite center of mass momentum $2\mathbf{q}$. Using this expression, we are now ready to calculate the superfluid weight. In the flat band limit, we can approximate $\epsilon(\mathbf{k})=\mu$, the resulting BdG quasiparticle energy is given as, $E_{\textrm{BdG}}(\mathbf{k})=\pm(\mu^2+|\Delta_{\textrm{proj}}(\mathbf{k},\mathbf{q})|^2)^{1/2}$. Then, the zero temperature superfluid weight is given by,
\begin{eqnarray}
[\mathbf{D}_s]_{i,j}=\frac{1}{V}\frac{\partial^2 F(\mathbf{q})}{\partial q_i \partial q_j}
=-\frac{1}{2V}\sum_{\mathbf{k}}
\frac{\partial_{i}\partial_j |\Delta_{\textrm{proj}}(\mathbf{k},\mathbf{q})|^2}{E_\textrm{BdG}(\mk)}
-\frac{\partial_{i}|\Delta_{\textrm{proj}}(\mathbf{k},\mathbf{q})|^2
	\partial_{j}|\Delta_{\textrm{proj}}(\mathbf{k},\mathbf{q})|^2}{2E_\textrm{BdG}(\mk)^{3}}
\label{Eq:sDs}
\end{eqnarray}

 To evaluate the above expression, we first need to calculate the $\mathbf{q}$ derivative of $\Delta_{\textrm{proj}}(\mk,\mathbf{q})$ (We use the \tcr{red} color to denote $\tcr{\mathbf{q}}$ and label the momentum just by a subscript for the clarity).
\begin{eqnarray}
|\Delta_{\textrm{proj}}(\mk,\tcr{\mathbf{q}})|^2
=
\sum_{N_{1,2,3,4}}
\big( [U^\dagger_{k+\tcr{q}}]_{n,N_1}[U_k]_{N_1,N}[U^T_{-k}]_{n,N_2} [U^*_{-k+\tcr{q}}]_{N_2,n}\big)
\big([U^T_{-k+\tcr{q}}]_{n,N_3}[U^*_{-k}]_{N_3,n}[U^\dagger_k]_{n,N_4}[U_{k+\tcr{q}}]_{N_4,n} \big)
|\Delta_{\textrm{proj}}(\mathbf{k})|^2.
\nonumber
\\
\end{eqnarray}
We find that the first order derivative vanishes. Therefore, the second order term vanishes in Eq. \ref{Eq:sDs}.
We calculate the second derivative of $\tcr{\mathbf{q}}$ as,
\begin{eqnarray}
\frac{\partial^2 |\Delta_{\textrm{proj}}(\mk,\tcr{\mathbf{q}})|^2}{\partial \tcr{q_i}\partial \tcr{q_j}}\bigg\rvert_{\tcr{\mathbf{q}}=0}=
(F_{4} +F_{12})|\Delta_{\textrm{proj}}(\mathbf{k})|^2,
\end{eqnarray}
which we decompose into the two parts. The first part, $F_{4}$, is given as,
\begin{eqnarray}
F_4=[\partial^2_{\tcr{i}\tcr{j}}U^\dagger_{\mk}U_\mk]_{n,n}
+[U^T_{-\mk}\partial^2_{\tcr{i}\tcr{j}}U^*_{-\mk}]_{n,n}
+[\partial^2_{\tcr{i}\tcr{j}}U^T_{-\mk}U^*_{-\mk}]_{n,n}
+[U^\dagger_\mk \partial^2_{\tcr{i}\tcr{j}}U_{\mk}]_{n,n},
\label{eq:SF4}
\end{eqnarray}
and the second part, $F_{12}$ is given as,
\begin{eqnarray}
F_{12}&=&
-[\partial_{\tcr{i}}U^\dagger_{\mk}U_\mk]_{n,n}[U^T_{-\mk} \partial_{\tcr{j}}U^*_{-\mk}]_{n,n}
-[\partial_{\tcr{i}}U^\dagger_{\mk}U_\mk]_{n,n}[\partial_{\tcr{j}}U^T_{-\mk}U^*_{-\mk}]_{n,n}
+[\partial_{\tcr{i}}U^\dagger_{\mk}U_\mk]_{n,n}[U^\dagger_k\partial_{\tcr{j}}U_{\mk}]_{n,n}
\\
\nonumber
&+&[U^T_{-\mk} \partial_{\tcr{i}}U^*_{-\mk}]_{n,n}[\partial_{\tcr{j}}U^T_{-\mk} U^*_{-\mk}]_{n,n}
-[U^T_{-\mk} \partial_{\tcr{i}}U^*_{-\mk}]_{n,n}[U^\dagger_\mk \partial_{\tcr{j}}U_{\mk}]_{n,n}
-[\partial_{\tcr{i}}U^T_{-\mk} U^*_{-\mk}]_{n,n}[U^\dagger_\mk \partial_{\tcr{j}}U_{\mk}]_{n,n}
\\
&+&(i\leftrightarrow j).
\nonumber
\end{eqnarray}
In the above expression, the first two terms and the last two terms cancel each other. The expression is further reduced to
\begin{eqnarray}
=[\partial_{\tcr{i}}U^\dagger_{\mk}U_\mk]_{n,n}[U^\dagger_k\partial_{\tcr{j}}U_{\mk}]_{n,n}
+[U^T_{-\mk} \partial_{\tcr{i}}U^*_{-\mk}]_{n,n}[\partial_{\tcr{j}}U^T_{-\mk} U^*_{-\mk}]_{n,n}
\\
\nonumber
+ [\partial_{\tcr{j}}U^\dagger_{\mk}U_\mk]_{n,n}[U^\dagger_k\partial_{\tcr{i}}U_{\mk}]_{n,n}
+[U^T_{-\mk} \partial_{\tcr{j}}U^*_{-\mk}]_{n,n}[\partial_{\tcr{i}}U^T_{-\mk} U^*_{-\mk}]_{n,n}.
\label{eq:SF12}
\end{eqnarray}
Summing up the contributions of Eq. \eqref{eq:SF4} and Eq. \eqref{eq:SF12}, we find that the total contribution is given by,
\begin{eqnarray}
\nonumber
\sum_{\mk}\frac{\partial^2 |\Delta_{\textrm{proj}}(\mk,\tcr{\mathbf{q}})|^2}{\partial \tcr{q_i}\partial \tcr{q_j}}\bigg\rvert_{\tcr{\mathbf{q}}=0}
&=&
-4\textrm{Re}\sum_{\mk} \big[\frac{1}{2}
([\partial_{\tcr{i}}U^\dagger_{\mk}\partial_{\tcr{j}}U_\mk]_{n,n}
+[\partial_{\tcr{j}}U^\dagger_{\mk}\partial_{\tcr{i}}U_\mk]_{n,n})
+[U^\dagger_{\mk} \partial_{\tcr{i}} U_\mk]_{n,n}[U^\dagger_k\partial_{\tcr{j}}U_{\mk}]_{n,n}\big]
|\Delta_{\textrm{proj}}(\mathbf{k})|^2
\\
&=&
-4\sum_{\mk}\textrm{Re} g_{ij}(\mathbf{k})|\Delta_{\textrm{proj}}(\mathbf{k})|^2,
\end{eqnarray}
In the last equality, we have used Eq. \eqref{eq:sReQ}. 
Accordingly, we find that 
\begin{eqnarray}
\nonumber
[\mathbf{D}_s]_{i,j}&=&
-\frac{1}{2V}\sum_{\mathbf{k}}
\frac{\partial_{i}\partial_j |\Delta_{\textrm{proj}}(\mathbf{k},\mathbf{q})|^2}{E_{BdG}(\mk)}
\\
&=&\frac{2}{V}\sum_{\mk}\frac{|\Delta_{\textrm{proj}}(\mathbf{k})|^2}{E_{\textrm{BdG}}(\mk)}\textrm{Re} g_{ij}(\mathbf{k}).
\end{eqnarray}
Finally, we arrive at Eq. \eqref{Eq:Ds2} in the main text.

\section{Geometric supercurrent}\label{sec:lg}
We consider the gradient coupling to the order parameter in the free energy of the Hofstadter superconductor. According to Eq. \eqref{Eq:Ds2} in the main text, we can write down the gradient term as (assuming no momentum dependence),
\begin{eqnarray}
f_{\textrm{grad}}\equiv F_{\textrm{grad}}/V
=\frac{1}{2\gamma} |(-i\hbar\nabla-\frac{2e}{c}\mathbf{A} )\Delta|^2.
\end{eqnarray}
where $\gamma\equiv \sqrt{\mu^2+|\Delta|^2}$. We now consider the variation of the free energy under the change of the gauge field, $\mathbf{A}\rightarrow \mathbf{A}+\delta \mathbf{A}$. It is given as,
\begin{eqnarray}
\frac{\delta f}{\delta \mathbf{A}} =\frac{\mathbf{j}_{\textrm{geo}}}{c}-\frac{e}{c\gamma}(\Delta(i\hbar\nabla-\frac{2e}{c}\mathbf{A})\Delta^*+\Delta^*(-i\hbar\nabla-\frac{2e}{c}\mathbf{A})\Delta)=0.
\end{eqnarray}
We derive the expression of the geometric supercurrent as,
\begin{eqnarray}
\mathbf{j}_{\textrm{geo}}=\frac{e|\Delta|^2}{\sqrt{\mu^2+|\Delta|^2}}  (\nabla \chi -\frac{2e}{c}\mathbf{A}).
\end{eqnarray}
which is Eq. \eqref{Eq:j} in the main text. We point out that the free energy is calculated in the zero temperature limit, and the argument does not rely on the basic assumption of the Landau-Ginzburg theory, which assumes the small order parameter amplitude. 


\end{widetext}

\end{document}